\documentclass[12pt,preprint]{aastex}

\def\beq{\begin{equation}}
\def\enq{\end{equation}}
\def\bea{\begin{array}}
\def\ena{\end{array}}

\def\msun{M_\odot}

\def\msun{$M_{\odot}$}

\def\mdot{$\dot M$}

\def\fdg{\hbox{$.\!\!^\circ$}}

\def\ergsec{\hbox{erg s$^{-1}$ }}

\def\fdg{\hbox{$.\!\!^\circ$}}

\bibpunct[,]{(}{)}{;}{a}{}{,}

\begin{document}

\title{~~A Determination of the Spin of the Black Hole Primary in
 \newline LMC X-1}

\author{ Lijun Gou, Jeffrey E. McClintock, Jifeng Liu, Ramesh Narayan,
  James F. Steiner}

\affil{Harvard-Smithsonian Center for Astrophysics, 60 Garden street,
Cambridge, MA 02138}

\author{Ronald A. Remillard}
\affil{Kavli Institute for Astrophysics and Space Research,
  Massachusetts Institute of Technology, Cambridge, MA 02139}

\author{Jerome A. Orosz}
\affil{Department of Astronomy, San Diego State University, 5500
  Companile Drive, San Diego, CA 92182}

\author{Shane W. Davis}
\affil{Institute For Advanced Study, Einstein Drive, Princeton, NJ 08540}

\author{Ken Ebisawa} \affil{Institute of Space and Astronautical
  Science/JAXA, 3-1-1 Yoshinodai, Sagamihara, Kanagawa 229-8510,
  Japan}

\author{Eric M. Schlegel}
\affil{Department of Physics and Astronomy, University of Texas at San Antonio,
  1 UTSA Circle, San Antonio, TX, 78249}

\begin{abstract}

  The first extragalactic X-ray binary, LMC X-1, was discovered in
  1969.  In the 1980s, its compact primary was established as the
  fourth dynamical black-hole candidate.  Recently, we published
  accurate values for the mass of the black hole and the orbital
  inclination angle of the binary system.  Building on these results,
  we have analyzed 53 X-ray spectra obtained by {\it RXTE} and, using
  a selected sample of 18 of these spectra, we have determined the
  dimensionless spin parameter of the black hole to be $a_* =
  0.92_{-0.07}^{+0.05}$.  This result takes into account all sources
  of observational and model-parameter uncertainties.  The standard
  deviation around the mean value of $a_*$ for these 18 X-ray spectra,
  which were obtained over a span of several years, is only $\Delta
  a_*=0.02$.  When we consider our complete sample of 53 {\it RXTE}
  spectra, we find a somewhat higher value of the spin parameter and a
  larger standard deviation. Finally, we show that our results based
  on {\it RXTE} data are confirmed by our analyses of selected X-ray
  spectra obtained by the {\it XMM-Newton}, {\it BeppoSAX} and {\it
    Ginga} missions.

\end{abstract}


\section{Introduction}

LMC X-1 was the first extragalactic X-ray binary to be discovered
\citep{mprs+69,pgrs+71}.  Its X-ray flux is quite constant, varying in
intensity by $\lesssim 25$\% during 12 years of monitoring by the
All-Sky Monitor (ASM) on board the {\it Rossi X-ray Timing Explorer}
\citep[{\it RXTE};][]{osmt+08}.  The source has been observed by
essentially all X-ray astronomy missions from {\it Uhuru} to {\it
Chandra} \citep[e.g., ][]{lkgt+71,cfzb+02}. Its bolometric, isotropic
luminosity is $L_{\rm bol} \approx 2.2 \times 10^{38}$~erg~s$^{-1}$,
which is $\approx 16$\% of its Eddington luminosity (\S5.1).

A dynamical black-hole model for the system was presented in pioneering
work by \citet{hcc+83,hcc+87}.  Recently, we have greatly improved the
dynamical model of the system \citep{osmt+08}.  The results of most
relevance to this paper are our determinations of the black hole mass $M
= 10.91 \pm 1.54$ \msun~and the orbital inclination angle $i = 36\fdg38 \pm
2\fdg02$ degree.  The distance is likewise important here; we adopt a
distance modulus of $18.41 \pm 0.10$ mag, which corresponds to $D =
48.10 \pm 2.22$ kpc \citep{omnb+07,osmt+08}.  Accurate values of $M$,
$i$ and $D$ such as these are crucial for the determination of black
hole spin by the X-ray continuum-fitting method that we employ
\citep{zcc+97,lznm+05}.

To date, our group has published spin estimates for four stellar-mass
black holes using the X-ray continuum-fitting method: GRO J1655--40,
$a_*=0.65-0.75$, and 4U 1543--47, $a_*=0.75-0.85$ \citep{smnd+06}; GRS
1915+105, $a_*=0.98-1.0$ \citep{msnr+06}; and M33 X-7,
$a_*=0.77\pm0.05$ \citep{lmnd+08}. The dimensionless spin parameter
$a_*\equiv cJ/GM^2$, where M and J are the mass and angular momentum
of the black hole; the value $a_*=0$ corresponds to a Schwarzschild
black hole and $a_*=1$ to an extreme Kerr hole. In this paper we show
that LMC X-1 harbors a rapidly spinning black hole.  Estimates of the
spins of stellar-mass black holes are also being obtained by modeling
the profile of the Fe K line (see
\citeauthor{mrfm+09}~\citeyear{mfwr+02,mfrn+04,mrfm+09}, and
references therein).

In all of our previous work cited above, we have exclusively used
thermal dominant (TD) spectral data in deriving our estimates of black
hole spin (\citealp{mr+06}; \citealp{rm+06}).  The definition of the
TD state is largely based on a spectral analysis that makes use of the
venerable multi-temperature and nonrelativistic disk model known as
{\sc diskbb} in XSPEC \citep{mikm+84,mmmb+86} plus the standard
power-law component {\sc powerlaw}~\citep{tl+95,mr+06,rm+06}; XSPEC is
a widely-used X-ray spectral fitting package \citep{arna+96}.
The key feature of the TD state is the strong presence of a soft ($kT
\sim 1$ keV) blackbody-like component of emission that arises in the
innermost region of the accretion disk.  The TD state is defined by
three criteria: (1) The ``disk fraction,'' the fraction of the total
2--20 keV unabsorbed flux in the thermal component, is $> 75$\%; (2)
the rms power in the power density spectrum integrated from 0.1--10 Hz
is $<0.075$; and (3) QPOs are absent or very weak (see Table
\ref{gold_spectra_result} and text in \citealp{rm+06}).

Following the selection of the TD spectral data, we determine the spin
parameter $a_*$ using our relativistic disk model {\sc kerrbb{\small
2}}~\citep{msnr+06}.  This model incorporates a lookup table for the
spectral hardening factor $f$ and returns two fit parameters, the spin
$a_*$ and the mass accretion rate \mdot.  In addition to the thermal
disk spectrum, there is also present a nonthermal tail component of
emission, which contributes a few percent of the 1--10 keV flux.  This
tail component is widely believed to originate from Compton upscattering
of the blackbody seed photons in a hot corona \citep{mr+06}.  Source
spectra obtained in the TD state are minimally corrupted by the
uncertain effects of Comptonization, and they are ideal for the
determination of spin via the continuum-fitting method.

In earlier work (see especially $\S$4.2 in \citealp{msnr+06}), we have
been hampered by the lack of an appropriate model of the tail
component of emission, which we have modeled as a standard power law
({\sc powerlaw} in XSPEC) or by using one of the existing
Comptonization models (e.g., {\sc comptt}).  We found both approaches
unsatisfactory because the power law diverges from realistic
Comptonization models at low energies (see \S4.2) while the existing
Comptonization models have more parameters than can be determined.
Very recently, we have developed a simple 2-parameter Comptonization
model in order to treat this tail component (\citealp{snme+08a};
\S\ref{simpl_model}).  The model, which is named {\sc simpl}, has been
implemented in XSPEC.  It is a convolution model that redirects a
fraction of the blackbody seed photons into a power law. The benefits
of using {\sc simpl} to model the tail component are that it
eliminates the unphysical divergence of a standard power law at low
energies while introducing just two parameters into the fit.  Its two
parameters are the scattered fraction $f_{\rm SC}$, the fraction of
the seed photons that are scattered into a power-law tail component,
and a photon index $\Gamma$.  An illustration of the difference
between the models {\sc powerlaw} and {\sc simpl} and a discussion of
the benefits of using the latter model are given in \S4.2.

In this paper, we focus on 53 {\it RXTE} spectra obtained using the
Proportional Counter Array (PCA).  These data are central because they
are abundant, and especially because they allow one to select data via
the rms variability criterion (Remillard \& McClintock 2006), which we
show is important for the reliable determination of the spin of LMC
X-1.  LMC X-1 was also observed in single observations using {\it
  XMM-Newton}, {\it BeppoSAX}, BBXRT, and {\it ASCA}, and on six
occasions by {\it Ginga}.  In Appendix \ref{other_mission_spin}, we
analyze the spectral data obtained by these five missions and find
values of spin that confirm our principal {\it RXTE} PCA results.

This paper is organized as follows. In \S2 we discuss the selection of
our gross sample of 53 {\it RXTE} spectra and their reduction.  In \S3
we present the results of a preliminary, nonrelativistic data analysis
and select a final sample of 18 spectra for analysis.  In \S4 we
discuss our relativistic data analysis with emphasis on our first use
of the Comptonization model {\sc simpl}.  We present our results in
\S5, a general discussion in \S6, and our conclusions in \S7.

\section{Data Selection and Reduction}

As noted in \S1, the data of central importance in this paper were
obtained using the large-area PCA detector onboard {\it RXTE}
\citep{swan+99}.  We consider 53 of the total of 55 PCA pointed
observations that have been performed during the course of the
mission; these 53 observations are identified and characterized in
Table 1.  We disregard two observations that were obtained very early
in the mission (1996 February 10 and 1996 March 8) because the
detector response is poorly known during that period, and the PCA
gain was higher than at later times.  The {\it RXTE} PCA data were
downloaded from NASA's High Energy Astrophysics Science Archive
Research Center (HEASARC).

In our analysis, we only include pulse-height spectra from PCU-2
because fits to the power-law spectrum of the Crab show that this is
the best calibrated of the five proportional counter units (PCUs) that
comprise the PCA.  Data reduction tools from HEASOFT version 5.2 were
used to screen the event files and spectra.  Data were taken in the
``Standard 2 mode,'' which provides coverage of the PCA bandpass every
16~s.  Data from all Xe gas layers of PCU-2 were combined to make the
spectra.  Background spectra were derived using the tool {\it
  pcabackest} and the latest ``faint source'' background model and
were then subtracted from the total spectra.  Redistribution matrix
files and ancillary response files were freshly generated individually
for each PCU layer and combined into a single response file using the
tool {\it pcarsp}.  In fitting each of the pulse-height spectra
(\S\S3,4), we used response files that were targeted to the time of
each observation of LMC X-1.

Following the methods spelled out in \S3 in \citet{msnr+06}, for the
53 observations of the source we individually (1) added the customary
systematic error of 1\% to all the PCA energy channels; (2) corrected
the PCA count rates for dead time; and (3) corrected the effective
area of the PCA using for each observation of LMC X-1 the proximate
and dead-time corrected observation of the Crab
Nebula~\citep{ts+74}. The dead-time correction factor for LMC X-1 is
$\approx 1.021$.  The effective-area correction factor is $\approx
1.076$, which requires us to apply a typical downward correction to
the fluxes of $\approx 0.930$.  The net effect of the dead-time and
effective-area corrections was to reduce the LMC X-1 fluxes that we
obtained from the analysis of each of our 53 spectra by $\approx
5$\%.

\section{Preliminary Data Analysis}

All of the data analysis and model fitting throughout the paper was
peformed using XSPEC version 12.4~\citep{arna+96}.  In our two earlier
papers that rely primarily on {\it RXTE} data to estimate the spins of
three black holes \citep{smnd+06,msnr+06}, for each source we
performed a preliminary analysis using the nonrelativistic disk model
{\sc diskbb} plus the standard power law {\sc powerlaw} for three
reasons: (1) to select the TD data \citep{rm+06}, (2) to explore the
data using this time-tested model, and (3) to compare our results with
published results.

For LMC X-1 we have likewise performed a nonrelativistic analysis
using this conventional model, which in XSPEC notation is expressed as
{\sc phabs(diskbb+powerlaw)} \citep{rm+06}, where {\sc phabs} is a
widely-used model of low-energy photoelectric absorption.  The spectra
were fitted over the energy range 2.5--20.0 keV, and the XSPEC command
``energies'' was invoked to accommodate the broader energy range
required by the convolution model {\sc simpl}. The hydrogen column
density was fixed at $N_{\rm H} = 4.6 \times 10^{21}$ cm$^{-2}$, as
determined by an analysis of {\it Chandra} HETG grating data
\citep{cfzb+02,osmt+08}.  The PCA fits are quite insensitive to the
choice of the lower cutoff energy because the low-energy absorption is
$\le 10$\% above 2.5 keV.  We obtained good fits to this basic model
for all 53 spectra without needing to add any of the customary
additional components (e.g., a Gaussian line or a broad Fe absorption
edge).  The fit results, along with some observational data (e.g.,
exposure times and count rates), are summarized in Table 1.  The two
pairs of parameters returned by the fit are the two {\sc diskbb}
parameters, the temperature at the inner edge of the accretion disk
$T_{\rm in}$ and a normalization constant $K$ (columns 7 \& 8), and
the two {\sc powerlaw} parameters, the photon index $\Gamma$ and a
normalization constant (columns 9 \& 10).

Columns 2 \& 3 in Table 1 identify the observations by calendar date
and the midpoint time of the observation.  Columns 4 \& 5 give the
total exposure time and the total source count rate.  The rms power in
column 6, which is a timing parameter, is of special interest, and we
return to it after commenting on the contents of the remaining columns
in the table.  The four spectral parameters described in the preceding
paragraph are listed in columns 7--10.  The disk fraction (DF) as
defined in \S1, which characterizes the relative strength of the
thermal component, is given in column 11.  The state of the source
(see Table 2 and text in \citealp{rm+06}) is given in column 12, and
the reduced chi-square and the number of degrees of freedom are given
in the final column.

An inspection of column 12 in Table 1 shows that only 3 out of our 53
spectra are classified as TD spectra (Nos. 12, 17 \& 18)\footnote{QPOs
  were absent in these spectra and all 53 {\it RXTE} PCA spectra,
  although they have been reported for LMC X-1 in {\it Ginga}
  observations by \citet{emi+89} and in {\it BeppoSAX} observations by
  \citet{dhgt+01}.}.  We consider our results for these spectra to be
of the highest reliability, and we therefore refer to them as our
``gold'' spectra.  However, this sample is small, and so we consider a
second tier of quality, namely those spectra that fail to meet the TD
criterion that the disk fraction exceed 75\%, but which do meet the TD
criterion that the rms power is $ < 0.075$ (\S1).  These are our 15
``silver'' spectra, and they correspond to the following observations
listed in Table 1: Obs.\ Nos. 1--11 and 13--16.  Our adopted value of
the spin parameter (\S5.1) is based on the results for all 18 gold and
silver spectra; in \S6 we show that our spin results for both types of
spectra are in excellent agreement.

Meanwhile, the 35 spectra in the lower part of Table 1 (Nos.\ 19--53)
fail to meet the first two TD selection criteria.  We refer to these
as our bronze spectra, and we do not include these data in determining
our adopted value of $a_*$.  However, in \S6 we show that these bronze
spectra yield a value of $a_*$ that is consistent with that obtained
from the gold/silver spectra, although this value is less well
constrained.  Throughout the paper we focus our attention on our
selected sample of 18 gold/silver spectra, all of which satisfy the
temporal TD criterion that the rms power is $<$ 0.075.

\section{Data Analysis}

The summary in Table 1 of our preliminary analysis results indicates
the challenges faced in determining the spin of LMC X-1 via the
continuum-fitting method.  All but three of the spectra fail to meet
the most fundamental criterion that defines the TD state, namely that
the unabsorbed thermal component comprise at least 75\% of the 2--20
keV flux (column 11).  A related problem is the steepness of the
power-law component ($\Gamma \gtrsim 3$; column 9), which causes the
power-law flux to rival or exceed the thermal flux at low energies, a
situation that we regard as unphysical for a Comptonization
model~\citep{snme+08a}.  Finally, 35 of the 53 spectra fail to meet a
second basic criterion of the TD state: rms continuum power $< 0.075$
(\S1; Table 1).

In this section, we describe in turn our workhorse relativistic disk
model {\sc kerrbb{\small 2}} and the Comptonization model {\sc simpl}.
We then summarize the complete set of parameters for the composite
model that we employ and close by discussing the necessity of fixing
the power-law index.  Meanwhile, in Appendix
\ref{contamination_pulsar}, we consider and reject the possibility
that the spectrum of LMC X-1 is contaminated by flux from the nearby
X-ray pulsar PSR B0540--69.

\subsection{The Relativistic Accretion Disk Model: KERRBB2} 

Our model of a thin accretion disk around a Kerr black hole
\citep{lznm+05} includes all relativistic effects such as frame
dragging, Doppler boosting, gravitational redshift and light bending.
It also includes self-irradiation of the disk (``returning radiation'')
and the effects of limb darkening.  The effects of spectral hardening
are incorporated into the basic model {\sc kerrbb} via a pair of look-up
tables for the spectral hardening factor $f$ corresponding to two
representative values of the viscosity parameter: $\alpha=0.01$ and 0.1
(see \S5.3).  The entries in this table were computed using a second
relativistic disk model {\sc bhspec} \citep{dbht+05,dh+06}.  We refer to
the model {\sc kerrbb} plus this table/subroutine as {\sc kerrbb{\small
2}}, which is the accretion-disk model that we use hereafter in this
paper.  The model {\sc kerrbb{\small 2}} has just two fit parameters,
namely the black hole spin $a_*$ and the mass accretion rate $\dot M$,
or equivalently, $a_*$ and the Eddington-scaled bolometric luminosity,
$L_{\rm bol}(a_*,\dot M)/L_{\rm Edd}$, where $L_{\rm Edd} = 1.3 \times
10^{38}(M/M_{\sun})$ \ergsec \citep{st+83}.  For a detailed discussion
of {\sc kerrbb{\small 2}} and its application, see \S4 in
\citet{msnr+06}.

Throughout the paper, we present detailed results only for the case
$\alpha=0.01$.  However, we have computed all of the same results for
the case $\alpha=0.1$ as well.  As we show in \S5.3, the choice of the
higher value of $\alpha$ results in a slightly lower value of $a_*$.
Importantly, our final adopted value for the spin parameter $a_*$ and
its error takes into account the inherent uncertainty in $\alpha$
(\S\S5.3,5.6).

\subsection{The Comptonization Model: SIMPL}
\label{simpl_model}

In this paper, we make the first application of the Comptonization
model {\sc simpl}~\citep{snme+08a}~in the determination of the spin of
a black hole.  The model was introduced in \S1 where our motivations
for developing it and its importance in this present work are
discussed.  The model is fully described in \citet{snme+08a} where it
is applied to data and its performance is shown to compare favorably
to such physical models of Comptonization as {\sc comptt} and {\sc
  compbb}.  The model {\sc simpl} (SIMple Power Law) functions as a
convolution that converts a fraction $f_{\rm SC}$ of seed photons into
a power law with photon index $\Gamma$.  There are two implementations
in XSPEC: {\sc simpl-{\small 1}}, which uses a one-sided Green's
function and corresponds to up-scattering only, and {\sc
  simpl-{\small2}}, which uses a classical two-sided Green's function
\citep{st+80} and corresponds to both up- and down-scattering of
photons.  {\sc simpl-{\small 1}} is a ``bare-bones'' and
computationally-efficient implementation.  In our application, the two
models give nearly identical results for the spin parameter.  Generically, throughout the paper we refer to both
versions of the code as {\sc simpl}; however, all of the analysis in
this paper is done using {\sc simpl-{\small 1}}.

Because {\sc simpl} is a convolution model it ties the Comptonized
component directly to the energy distribution of the input photons,
which in the application at hand is specified by our disk model {\sc
kerrbb{\small 2}}.  The most important feature of {\sc simpl} is that
it produces a power-law tail at energies higher than those of the
thermal seed photons, but the power law does not extend to lower
energies, which is a property shared by all physical Comptonization
models (e.g., {\sc comptt}).  Thus, the crucial difference between
{\sc simpl} and the standard power law is that {\sc simpl} cuts off in
a physically natural way, whereas the standard power law diverges at
low energy.  At the same time, {\sc simpl} is an empirical model that,
while mimicking the mechanics of scattering, makes as few assumptions
as possible.  In particular, the model has only two free parameters,
as few as the standard power law.

Figure 1 starkly illustrates the quite different models one obtains
employing {\sc powerlaw} versus {\sc simpl}.  Shown are fits to one of
our 18 selected spectra using the models {\sc phabs(diskbb+powerlaw)}
and {\sc phabs(simpl$\otimes$diskbb)}.  Contrasting behaviors between
{\sc powerlaw} and {\sc simpl} are most clearly revealed at low
energies in the unabsorbed models, which are shown in the rightward
pair of panels.  Note how the power-law component rises without limit,
whereas the {\sc simpl} model cuts off in a physically natural way in
the manner of all Comptonization models.  Because {\sc powerlaw}
produces higher fluxes than {\sc simpl} at low energies, it steals
flux from the thermal component and hardens it, thereby reducing its
normalization constant.  Furthermore, the rising power-law component
yields inflated estimates of $N_{\rm H}$ relative to {\sc simpl} and
other Comptonization models.  Thus, although the quality of fit using
either of these two-parameter models is comparable, they can yield
quite different results for both the parameters returned by the
thermal component and for $N_{\rm H}$.  These differences between {\sc
  powerlaw} and {\sc simpl} are most pronounced when the power law is
relatively steep, $\Gamma \gtrsim 3$, as in the case of LMC X-1.

When the model {\sc simpl} is convolved with a thermal component, in
addition to the observed thermal component, one also has available the
seed spectrum, i.e., the thermal spectrum prior to Comptonization.
Thus, there are two possible definitions of the disk fraction: the
classical definition of the disk fraction DF that is based on the
observed component of thermal flux (Remillard \& McClintock 2006), and
a new definition based on the flux of the blackbody seed photons,
i.e., the model-dependent flux that emerges from the surface of the
disk ingoring the effects of the corona that dresses it.  We refer to
this disk fraction as the ``naked disk fraction'' or $NDF$.

The two disk fractions are simply related via the scattered fraction
$f_{\rm SC}$: $DF = (1-f_{\rm SC})NDF$.  The choice of which of the
two disk fractions to use depends on the purpose at hand.  The disk
fraction $DF$ is currently used to classify the state of the source
(\S3), whereas the naked disk fraction $NDF$ is useful for assessing
energetics (e.g., in relating the X-ray power to the thermal energy
stored in the corona).  Throughout this paper, we have chosen to use
the naked disk fraction and we compute it for the canonical energy
interval 2--20 keV.


\subsection{Description of Model Parameters}

In all of the relativistic data analysis, we use {\sc kerrbb{\small
    2}} convolved with {\sc simpl} in conjunction with {\sc phabs},
which models the low-energy absorption: i.e., our XSPEC model is {\sc
  phabs(simpl}$\otimes${\sc kerrbb{\small 2}}).  We use precisely the
same energy range, 2.5--20.0 keV, and fixed hydrogen column density,
$N_{\rm H} = 4.6 \times 10^{21}$ cm$^{-2}$, that we used in our
preliminary data analysis (\S3).  The number of fit parameters is
likewise the same, namely four in total: two for {\sc kerrbb{\small
    2}}, the spin parameter $a_*$ and the mass accretion rate \mdot,
and two for {\sc simpl}, the power-law index $\Gamma$ and the
scattered fraction $f_{\rm SC}$.  The normalization constant of {\sc
  kerrbb{\small 2}} is fixed at unity, which is appropriate when $M$,
$i$ and $D$ are held fixed at the values quoted in \S1.  For {\sc
  kerrbb{\small 2}} we also include the effects of limb darkening
(lflag=1) and returning radiation (rflag=1), and the torque at the
inner boundary of the accretion disk is set at zero ($\eta=0$;
\citealp{smnt+08,snm+08}).  In the case of the Comptonization model
{\sc simpl}, we set switch UpScOnly=1, i.e., we use the {\sc
  simpl-{\small 1}} version of the code, which corresponds to
up-scattering only (\S4.2).

\subsection{Fixing the Power-Law Index $\Gamma$}

We further reduced the number of free fit parameters from four to three
by fixing the spectral index $\Gamma$.  We did this because the
power-law component is not well constrained, as we now explain.  LMC X-1
is classified as a faint source for the PCA, i.e., its total count rate
is $< 40$ counts cm$^{-2}$ s$^{-1}$ per PCU (Table 1).  The PCA
background rate is determined using a model, in this case the ``faint
source background model,'' which is keyed to the count rate of a
charged-particle detector aboard the {\it RXTE} spacecraft
\citep{jmrr+06}.

We find that there are quite significant systematic uncertainties in
the background count-rate model.  This is clearly shown by computing
the count rates for the 18 gold/silver PHA spectra for the energy
range 20--40 keV, where our models predict that the source count rate
is completely negligible relative to the background rate. We first
note that these background rates indicate that the model background
has been overestimated because for 14 of the 18 spectra the rates are
negative.  More telling is the fact that these 18 background rates
deviate from the expected value of zero count rate on average by $4.6
\sigma$ (std.\ dev; N = 18). These significant deviations show that
the background model provides a poor approximation to the true
background rate.

We now consider the net source count rates for the 18 gold/silver PHA
spectra for the energy range 10--20 keV, which is crucial for the
determination of the power-law parameters (while being completely
unaffected by the $\sim 1$~keV thermal component).  Using the 20--40
keV photon rates as a proxy for the 10--20 keV background rates, we
find for half of the 18 observations (Nos.\ 1--3, 7--9 and 15--17 in
Table 1) that the background rates are fully one-fifth of the source
count rates.  Subtracting this relatively large and uncertain
background rate from the total count rate significantly corrupts the
source spectrum.  Thus, for this faint source we conclude that we
cannot adequately constrain the power-law index, and we therefore fix
it at the reasonable nominal value of $\Gamma=2.5$ \citep{rm+06}.  The
good news is that our final value of the spin parameter $a_*$ is very
insensitive to this choice of $\Gamma$, as we show in \S5.2.

\section{Results}

We first present our featured results for the 18 gold/silver spectra
in \S\ref{primary_results}.  In \S\ref{different_model_results} these
results are compared to those obtained using the standard power law in
place of {\sc simpl} and also to those obtained using different fixed
values of $\Gamma$.  In \S5.3 we examine the sensitivity of our
results to varying the metallicity $Z$ and the viscosity parameter
$\alpha$.  In \S\ref{all_spectra_results} we consider all 53
gold/silver/bronze spectra and examine the correlations between
selected pairs of the fit parameters.  Then, in
\S\ref{new_disk_fraction} we show how the scattered fraction and the
naked disk fraction depend systematically on the rms power, and we
explore the usefulness of the scattered fraction $f_{\rm SC}$ in
selecting data.  Finally, in \S\ref{error_analysis} we derive the
error in the spin parameter $a_*$ that includes all sources of
observational uncertainty, as well as the uncertainties in the model
parameters.  In \S\S5.1--5.5, the parameters $M,~i$, and $D$ are held
fixed at their best central values (see \S1), and they are allowed to
vary only in \S5.6.

\subsection{Primary Results}
\label{primary_results}

Our featured result for the spin parameter is given in the bottom line
of Table 2: $a_* = 0.938 \pm 0.020$ (std.\ dev.; $N=18$).  Note that
this value, which is based on all 18 gold/silver spectra, is entirely
consistent with the value obtained by analyzing just the three gold TD
spectra ($a_*=0.928\pm0.014$; see footnote $c$ in Table 2).

Table 2 contains all of our fit results for the 18 gold/silver spectra
obtained using our relativistic disk model convolved with our
Comptonization model.  The uncertainties for the mean values given in
the bottom line are the standard deviation for $N=18$.  From left to
right, $a_*$ and \mdot~are the fitted values of the spin parameter and
the mass accretion rate, and $f$ is the interpolated value of the
spectral hardening factor (\S4.1).  The two fit parameters of {\sc
  simpl} follow: the photon index $\Gamma$ (which is held fixed at
2.5) and the scattered fraction $f_{\rm SC}$~(which is fitted).  The
next quantity is the naked disk fraction as defined for our
Comptonization model (see \S4.2).  In the following column, the value
of $\chi_{\nu}^{2}$ and the number of degrees of freedom show that the
fits are all quite good with $\chi^2_{\nu} < 1.4$.  The final column
gives the Eddington-scaled bolometric luminosity (\S4.1) corresponding
to the naked disk, which is larger than the observed disk luminosity
by the factor $1/(1-f_{\rm SC})$, i.e., by 3-5\%.  For all the results
given in Table 2 the photon index was fixed at $\Gamma=2.5$ (\S4.4),
and the column density was fixed at $N_{\rm H} = 4.6 \times 10^{21}$
cm$^{-2}$ (\S3).

The luminosity is a modest fraction of the Eddington value and shows
little variation over the 1996--2004 period during which these 18
spectra were obtained: $L/L_{\rm Edd} = 0.160 \pm 0.007$ (Table 2).
The relative constancy of LMC X-1 is also attested to by the 12-year
record of the {\it RXTE} All-Sky Monitor: The intensity of LMC X-1 has
been more or less steady at $\approx 20$ mCrab (1.5--12 keV) with
variations averaged over 10-day intervals of only $\approx \pm25$\%
\citep{osmt+08}.  In this work, as in all previous work, we select
only those observations for which $L/L_{\rm Edd} < 0.3$, which
corresponds to a disk that is geometrically thin at all radii, with a
height-to-radius ratio less than 0.1 \citep{msnr+06}.  An inspection
of the last column in Table 2 shows that all 18 gold/silver spectra
easily meet this requirement.

On a time scale of many thousands of years, one fully expects the spin
of the black hole to be constant, essentially unaffected by either
accretion torques or any process that might extract angular momentum
\citep{kk+99,lmnd+08}.  As shown in
Figure~\ref{evolution_spin_parameter} and as expected, the spin of LMC
X-1 over the 8-year span of the {\it RXTE} observations is consistent
with a constant value of $a_*$ to within $\approx 2$\% (std.\ dev.;
N=18).

Importantly, the high spin we find using {\it RXTE} PCA data, $a_* =
0.938 \pm 0.020$, is confirmed by X-ray missions with low-energy
response that reaches well below the 2.5 keV cutoff of the PCA. As we
show in Appendix A, when we apply the data selection criterion
developed in \S5.5 ($f_{\rm SC} < 4\%)$, we find values of spin using
spectra obtained by the {\it XMM-Newton}, {\it BeppoSAX}, and {\it
  Ginga} missions that are consistent with those we obtain using our
{\it RXTE} spectra.

\subsection{Results Obtained Using Other Models and Using Different 
Values of $\Gamma$}
\label{different_model_results}

In all earlier work, we have relied on the standard model {\sc
  phabs(kerrbb{\small 2}+powerlaw)} in deriving the spin parameter
(\S3).  We found that this model was generally satisfactory (but see
\S4 in \citealp{msnr+06}) because the power law was not too steep
($\Gamma \sim 2.5$).  We now apply this old model to the 18
gold/silver spectra of LMC X-1 with $\Gamma \gtrsim 3$ (Table 1) in
order to see how much the steepness of the power law affects our
determination of $a_*$.  In making this comparison, we follow exactly
the steps in our earlier analysis except that we model the Comptonized
tail component of emission using the additive component {\sc powerlaw}
rather than convolving {\sc kerrbb{\small 2}} with {\sc simpl}.  For
example, we fix $\Gamma=2.5$ and $N_{\rm H} = 4.6 \times
10^{21}$~cm$^{-2}$ and fit over the energy range 2.5--20.0 keV.  The
resultant mean fit parameters for the 18 spectra are given in line 3
of Table 3.  For ease of comparison our featured mean fit parameters,
which are repeated from Table 2, are given in the first line of Table
3.  Comparing lines 1 \& 3, one sees that the old model yields a value
of $a_* = 0.953 \pm 0.018$, which differs from our featured value of
$a_* = 0.938 \pm 0.020$ by only 0.7 standard deviation.

How sensitive are our adopted results in Table 2 to our chosen value
of $\Gamma=2.5$ (\S4.4)?  In order to answer this question, we have
recomputed the results given in Table 2 with the photon index fixed at
$\Gamma=2.3$ and $\Gamma=2.7$, and we find respectively, $a_* = 0.944
\pm 0.019$ and $a_* = 0.932 \pm 0.021$.  These values differ from our
adopted value of $a_* = 0.938 \pm 0.020$ by only 0.2 standard
deviations.  Even if we fix the photon index at $\Gamma=3.1$, the mean
value in Table 1 for the 18 gold/silver spectra (as well as for all 53
spectra), we find $a_* = 0.918 \pm 0.024$, which differs from our
adopted value by 0.6 standard deviation.  This small difference is not
unexpected given that so few thermal photons are scattered into the
tail component, $f_{\rm SC} = 0.036 \pm 0.005$ (Table 2).  That $a_*$
is anti-correlated with $\Gamma$ is expected because a steeper
power-law component steals photons from the high-energy tail of the
thermal component, and the resultant softening of this component
implies a lower value of spin (and vice versa).

As a final detail, we note that for just 4 of our 18 spectra there is
clear evidence for an Fe emission line in the fit residuals.  In
deriving our primary results presented in \S5.1, we did not include an
Fe line because the fits are quite good without it.  As a test, we
have refitted all 18 spectra while including the Fe line, and we find
a resulting mean spin parameter of $a_*=0.922\pm0.021$, which is quite
consistent with our featured result above (see footnote $e$ in
Table~2).

\subsection{Results for Different Values of Metallicity and Viscosity 
Parameter}
\label{different_para_results}

The metallicity of the LMC, and presumably LMC X-1 as well, is $Z
\approx 0.3Z_{\odot}$ \citep{ds+03}, whereas our results are all given
for $Z = Z_{\odot}$.  How important is this difference in metallicity?
We recalculated all of the results summarized in Table 2 for the 18
gold/silver spectra using readily-available {\sc BHSPEC} table models
that were computed for $Z = 0.1Z_{\odot}$.  We found $a_* =
0.937\pm0.020$ (std.\ dev.; N = 18), which is almost identical to the
bottom-line result quoted in Table 2 ($a_* = 0.938\pm0.020$).  Thus,
varying the metallicity has a completely negligible effect on our
results.

In order to streamline the paper, throughout we quote detailed results
only for one value of the viscosity parameter, $\alpha=0.01$.
Presently, there is an inherent uncertainty in $\alpha$, and we direct
the reader to \citet{kpl+07} for a more comprehensive review of
numerical and observational constraints.  In addition, a detailed
discussion of {\sc BHSPEC}'s spectral dependence on $\alpha$ can be
found in \citet{dd+08}.  A low value is motivated by shearing box
simulations with zero net magnetic flux, which show $\alpha$ approaching
zero as resolution increases \citep{pcp+07,fp+07}.  The spectral models
are remarkably insensitive to $\alpha$ for $\alpha\le0.01$.  Therefore,
results obtained for models with $\alpha=0.01$ are representative of
those obtained for models with even lower values of $\alpha$.

Meanwhile, larger values of the viscosity parameter, $\alpha\sim0.1$,
are suggested by global GRMHD simulations \citep{hk+01,smnt+08} and
the disk instability model \citep{las+01}.  Our results for $a_*$ are
modestly sensitive to these larger values of $\alpha$.  Therefore, we
have recalculated all of the results summarized in Table 2 using {\sc
  BHSPEC} table models computed for $\alpha = 0.1$.  In this case, we
find a slightly lower value of the spin parameter:
$a_*=0.908\pm0.023$.  This value is 1.0 standard deviations less than
the value given in the bottom line of Table 2 for $\alpha=0.01$ ($a_*
= 0.938\pm0.020$).  In deriving our final adopted value of the spin
parameter $a_*$ and its error (\S5.6), we take into account the
uncertainty in the viscosity parameter by computing our results using
both of our fiducial values of $\alpha$: 0.01 and 0.1.

\subsection{Graphical Presentation of Results for All 53 Spectra}
\label{all_spectra_results}

Our adopted results (\S5.1) are based exclusively on the 18
gold/silver spectra.  All of the fitting results for these spectra are
given in Table 2, and histograms for the four fit parameters are shown
as filled blocks in Figure \ref{rxte_histogram}.  Although we do not
present a table of our results for the 35 bronze spectra (Nos. 19--53
in Table 1), we do show histograms based on the results of our
analysis of these spectra as open blocks in Figure
\ref{rxte_histogram}, which are superposed on the solid-block
histograms.  An inspection of the figure shows that, when compared to
the 18 gold/silver spectra, the fits to all 53 spectra yield on
average higher values of the spin parameter $a_*$, lower mass
accretion rates (\mdot), more scattering of the thermal photons
($f_{\rm SC}$) and smaller naked disk fractions ($NDF$s).  A
quantitative comparison of these differences is provided by a
comparison of the results for the 18 gold/silver spectra, which are
given in the top line of Table 3, with the comparable results we
obtained for all 53 spectra, which are given in the bottom line of
Table 3.

Figure \ref{rxte_correlation} shows correlation plots between pairs of
fit parameters.  The 18 gold/silver spectra are indicated by the filled
circles and the 35 bronze spectra by open circles.  Figure
\ref{rxte_correlation}$a$ shows a strong correlation between $a_*$ and
the mass accretion rate \mdot.  The range of the correlation is much
larger for the bronze spectra, extending to $a_* \approx 1$.  This
correlation is rather uninteresting because it simply reflects the fact
that \mdot~is a normalization constant that must necessarily decrease if
the measured parameter $a_*$ fluctuates upward (or vice versa).  In
Figure \ref{rxte_correlation}$b$ we recast this correlation by plotting
$a_*$ versus $\eta$\mdot, where the efficiency parameter $\eta = L_{\rm
bol}/\dot M{c^2}$ is computed for the particular value of $a_*$.  As
shown, there is no correlation between $a_*$ and $\eta$\mdot, i.e.,
between $a_*$ and the bolometric luminosity.

Figure \ref{rxte_correlation}$c$ shows a tight correlation between the
naked disk fraction and the scattered fraction.  Our adopted disk
fraction (\S4.2) is the ratio of the flux in the naked disk component
$F_{\rm ND}$ to the total unabsorbed flux from 2--20 keV, which is the
observed thermal flux plus the power-law component, $(1-f_{\rm
  SC})F_{\rm ND}$+$F_{\rm PL}$. As $f_{\rm SC}$ increases and thermal
photons are directed into the power-law tail, $F_{\rm PL}$ is expected
to increase substantially because each Compton-upscattered photon is
boosted in energy.  Thus, as $f_{\rm SC}$ and $F_{\rm PL}$ increase,
the disk fraction must decrease, as shown.

The relationship between $a_*$ and the scattered fraction is shown in
Figure 4$d$.  There is no clear correlation for the 18 gold/silver
spectra.  However, among the 35 bronze spectra $a_*$ is seen to
increase with scattered fraction, although the correlation is weak and
there are two discrepantly low points.

\subsection{Using the Scattered Fraction to Select Data}
\label{new_disk_fraction}


Figure \ref{rxte_rms_sf} shows correlation plots of the scattered
fraction and naked disk fraction versus the rms power (\S1).  The
solid vertical line at rms power = 0.075 divides the bronze data on
the right from the gold/silver data on the left.  These latter data
are weakly correlated, showing a tendency for the scattered fraction
to increase and the naked disk fraction to decrease as the rms power
increases.  To the right of the dividing line the sense of the
correlations are the same, but the correlations steepen and become
very pronounced.  Of course, one expects the scattered fraction and
the naked disk fraction to correlate in the opposite sense given the
tight anti-correlation between these quantities that is shown in
Figure \ref{rxte_correlation}$c$.



The TD-state criterion that the rms power is $<0.075$ has played a
central role in the selection of our data (\S3).  Unfortunately, the
use of the rms power criterion, which requires timing data up to $\sim
10$ Hz and is based on the particular broad energy response of the
PCA, is applicable to {\it RXTE} data only.  Meanwhile, the good
correlation between the rms power and the scattered fraction
(Figure~\ref{rxte_rms_sf}) allows the latter quantity to be used in
selecting data as a proxy for the rms power. (One could use the naked
disk fraction instead of $f_{\rm SC}$; however, we favor $f_{\rm SC}$
because it is a readily-available fit parameter.)  We now investigate
this possibility in the case of LMC X-1 using the model {\sc
  phabs(simpl}$\otimes${\sc kerrbb{\small 2})}.

In the top panel of Figure \ref{rxte_rms_df_corr}, we show the rms
power for all 53 spectra plotted in order of increasing rms power.
The horizontal line marks the TD-state criterion rms $< 0.075$ and the
vertical line separates our 18 gold/silver spectra on the left from
the 35 bronze spectra on the right.  With the same ordering of the
spectra, we show in the lower panel for all 53 spectra the scattered
fraction.  The dashed line for $f_{\rm SC}=4$\% was determined as
follows: We assumed that both the rms power and $f_{\rm SC}$ follow
normal distributions and are uncorrelated with the other fitted
parameters.  For the rms power we find a mean value of rms~$= 0.081
\pm 0.016$ (std.\ dev.; $n=53$), which is $0.38 \sigma$ above the rms
$=0.075$ line in the top panel.  Given the correlation between $f_{\rm
  sc}$ and rms power (Figure \ref{rxte_rms_sf}), we expect the
corresponding scattered-fraction cut to also be $0.38 \sigma$ below
the mean value, i.e., at $f_{\rm sc} = 4$\% (rounded to the nearest
percent); this is the level indicated by the dashed line in the lower
panel.

The solid symbols in the top panel correspond to the 16 spectra in the
lower panel that have $f_{\rm sc} < 4$\%.  It can be seen that all
of the spectra selected by scattered fraction are consistent or nearly
consistent with the TD criterion rms $ < 0.075$.  Likewise, in the
lower panel one sees that most (14 out of 18) of the gold/silver
spectra have $f_{\rm sc} < 4$\% while 4 of them have $4\% < f_{\rm sc}
\lesssim 5$\%.  Meanwhile, most of the 35 bronze spectra with rms
significantly greater than 0.075 are soundly rejected by the criterion
$f_{\rm sc} < 4$\%.

We now compare the fitting results obtained for the data selected
using the criterion rms $< 0.075$ to that selected using the criterion
$f_{\rm sc} < 4$\% and find that they are essentially identical: The
spin parameter and the scattered fraction for the rms-selected data
are respectively $a_*=0.938\pm0.020$ and $f_{\rm sc}=0.036\pm0.005$
(Table 2), and for the $f_{\rm SC}$-selected data are
$a_*=0.938\pm0.016$ and $f_{\rm SC}=0.033\pm0.003$. These results are
based only on LMC X-1.  However, they indicate that whenever
high-quality timing data is unavailable for a source that it may be
possible to establish a useful scattered-fraction criterion capable of
rejecting data with high flicker noise.  Of course, the
scattered-fraction cut level that one adopts may vary from source to
source.

In Appendix A, we make the first use of our newly-defined scattered
fraction criterion, $f_{\rm SC} < 4$\%, in order to select data
obtained by five X-ray missions for which the {\it RXTE} rms
data-selection criterion cannot be applied.


\subsection{Comprehensive Error Analysis: Final Determination of Spin
Parameter}

\label{error_analysis}

The statistical error in the spin parameter is quite small: $a_* =
0.938 \pm 0.020$ ($\alpha=0.01$; Table 2).  At this level of
precision, the observational error is dominated by the uncertainties
in the input parameters: the black hole mass $M$, the orbital
inclination angle $i$ and the distance $D$ (\S1;
\citealp{osmt+08}). In order to determine the error in $a_*$ due to
the combined uncertainties in $M$, $i$ and $D$, we performed Monte
Carlo simulations assuming that the uncertainties in these parameters
are normally and independently distributed.  Specifically, we (1)
generated 3000 parameter sets for $M$, $i$ and $D$; (2) computed for
each set the look-up table for the spectral hardening factor $f$ using
the model {\sc bhspec}; and (3) using these $f$-tables, we obtained
$a_*$ by fitting our standard model (\S4.3) to the 18 gold/silver
spectra.

The results for the 3000 simulations are shown in Figure
\ref{rxte_error_analysis}.  Panels $a$--$c$ show the distribution of
the values of $a_*$ for the parameters $M$, $i$ and $D$,
respectively. Each set of 18 spectra is averaged and represented by a
single data point.  In each panel, the value of the parameter in
question is given \citep{osmt+08}, and a large central dot is plotted
among the 3000 distributed points at $a_* = 0.937$, which is the
median value of the distribution. ~The histogram for the displacements
in $a_*$ about the most probable value is shown in panel $d$.  As
indicated, the $1~\sigma$ error in $a_*$ is (+0.040, -0.061), which is
much larger than the statistical error of $\pm 0.020$ (Table 2).  The
error is dominated by the uncertainty in $M$; the uncertainties in $i$
and $D$ are relatively unimportant.

The analysis above is for our standard value of $\alpha=0.01$.  We
also performed the Monte Carlo analysis for our other fiducial value
of the viscosity parameter, $\alpha=0.1$, which as we saw in \S5.3
results in a slightly lower mean value of the spin parameter, $a_* =
0.908\pm0.023$.~(The computed median value of the distribution is
$a_*=0.907$.)~ In Figure \ref{different_viscosity}, we show the
results of this analysis as a dashed-line histogram of spin
displacements; the histogram for $\alpha=0.01$, which is copied from
Figure \ref{rxte_error_analysis}$d$, is also shown by the smaller
solid-line histogram.  The summation of these two histograms results
in the large histogram (heavy solid line).  This combined
  distribution, which corresponds to a total of 6000 simulations, has
  a median spin value of 0.92 and implies a $1\sigma$ error of
(+0.05, -0.07).  {\it Thus, considering both the uncertainties in $M$,
  $i$ and $D$ and the inherent uncertainty in the value of the
  viscosity parameter we arrive at our final result for the spin of
  LMC X-1: $a_* = 0.92_{-0.07}^{+0.05}$} ($1 \sigma$).

\section{Discussion}
\label{discussion}

Throughout, we have focused on the three gold spectra, which fully
satisfy the TD-state criteria, and 15 additional silver spectra.
These latter spectra meet the TD-state rms power criterion. However,
they yield low values of the disk fraction (Table 1) that fail to meet
the TD-state disk-fraction criterion because the standard power-law
model is inadequate for modeling their steep Compton-tail components.
We have shown that these 18 gold/silver spectra when analyzed using
the Comptonization model {\sc simpl} yield large values of the naked
disk fraction (Table 2).  Restricting ourselves to this sample of 18
spectra and considering only the statistical error, we obtained our
featured value of the spin parameter: $a_* = 0.938 \pm 0.020$ (std.\
dev.; N=18).  For comparison, if we consider all 53 spectra (i.e.,
including the bronze spectra) we obtain: $a_* = 0.961 \pm 0.033$
(std.\ dev.; N=53; Table 3).  That is, we obtain a statistically
consistent result, although with a somewhat higher spin and with a
larger uncertainty.  We have chosen to base our result on just the 18
gold/silver spectra because they meet the TD-state rms criterion and
because their spectra are strongly dominated by the thermal component,
i.e., their 2--20 keV naked disk fractions are $\gtrsim 86$\% when
computed using {\sc simpl} (Table 2). When we include all the sources
of observational error and model parameter uncertainties, our final
result for the spin parameter is $a_*=0.92^{+0.05}_{-0.07}$.

The greatest uncertainty in our spin estimate arises from the
uncertainties in the validity of the disk model we employ.  For
example, the model assumes a razor-thin disk for which the viscous
torque is assumed to vanish at the innermost stable circular orbit
(ISCO), an assumption that, it has been argued, does not hold in the
presence of magnetic fields \citep{kh+02}.  However, our preliminary
hydrodynamic work indicates that the effects of these viscous torques
and emission from within the ISCO are modest for the thin disks
($H/R\le0.1$) and low luminosities ($L/L_{\rm Edd}\le0.3$) that we
restrict ourselves to \citep{msnr+06,smnd+06}.  In the case of LMC
X-1, with its modest luminosity ($L/L_{\rm Edd} \approx 0.16$; Table
2) and high spin, these hydrodynamic results imply a very small
uncertainty, $\Delta a_* \lesssim 0.01$ (see Fig.\ 9 in
\citealp{snm+08}).  Another uncertainty lies in our assumption that
the black hole spin is approximately aligned with the angular momentum
vector of the binary.  As Figure 3$b$ indicates, if any misalignment
is $\lesssim 2^{\rm o}$, then it will contribute an error in the spin
parameter $a_*$ that is no larger than our total observational error.
A future X-ray polarimetry mission may allow a check on our assumption
of alignment \citep{lnm+08}.

Our preliminary hydrodynamic results are supported by our more recent
three-dimensional general relativistic MHD simulations of an
appropriately thin disk with $H/R \sim 0.05-0.1$ \citep{smnt+08}.
This model is for a non-spinning black hole.  In steady state, the
specific angular momentum profile of the inflowing magnetized gas
deviates by $<2$\% from that of the standard thin disk model of
\citet{nt+73}.  In addition, the magnetic torque at the ISCO is only
$\sim2$\% of the inward flux of angular momentum at this radius.
These results indicate that magnetic coupling across the ISCO is
relatively unimportant for geometrically thin disks, at least in the
case of a non-spinning black hole.

In short, our recent theoretical studies indicate that the disk
emission is truncated rather sharply at the ISCO and that measuring
the truncation radius provides a reliable estimate of spin.  In
addition, there is a long history of observational evidence suggesting
that fitting the X-ray continuum is a promising approach to measuring
black hole spin.  This history begins in the mid-1980s with the simple
non-relativistic multicolor disk model \citep{mikm+84,mmmb+86} now
known as {\sc diskbb}, which returns the color temperature $T_{\rm
  in}$ at the inner-disk radius $R_{\rm in}$ (\S3).  In their review
paper on black hole binaries, \citet{tl+95} summarize examples of the
steady decay (by factors of 10--100) of the thermal flux of transient
sources during which $R_{\rm in}$ remains quite constant (see their
Fig.\ 3.14).  They remark that the constancy of $R_{\rm in}$ suggests
that this fit parameter is related to the radius of the
ISCO. \citet{zcc+97} then outlined how, using a relativistic disk
model and corrections for the effects of radiative transfer, the fixed
inner disk radius provides an observational basis to infer black hole
spin.  More recently, this evidence for a constant inner radius in the
thermal state has been presented for a number of sources via plots
showing that the bolometric luminosity of the thermal component is
approximately proportional to $T_{\rm in}^4$
\citep{kme+01,km+04,gd+04,afkk+05,mns+07}.

We now consider the possibility that our results are affected by a
component of absorption that is not included in our
model. \citet{cbsk+08} have found evidence for an ionization cone
within a few parsecs of LMC X-1.  They argue that the cone must be
powered primarily by photoionzation from the accretion disk, but
require a 13--300 eV photon flux which exceeds the extrapolated hard
X-ray model by a factor $\sim10$.  Both this excess and the anisotropy
of the ionized gas suggest that the X-ray emission might be partially
obscured near the source, possibly by the outer disk.  If this
hypothesis is correct, we could be underestimating the true disk
luminosity and, therefore, overestimating the spin.  However, we stand
by our estimate that the bolometric luminosity of the disk at the
source is $L/L_{\rm Edd} \sim 0.16$ (Table~2) while regarding the
evidence that the disk is substantially more luminous as tentative for
several reasons: e.g., the possibility of accretion rate variations
over the recombination time of the ionized gas; the lack of evidence
for obscuration in the X-ray spectrum; the close proximity of other
external ionizing sources; and the likelihood of significant EUV flux
via X-ray reprocessing in the accretion disk.  On this last point, our
quantitative estimate indicates that reprocessing can provide the
ionizing photon flux required to explain the He~II~$\lambda$4686 line
flux: Using the accretion-disk reprocessing model described in
Appendix A of \citet{osmt+08}, and assuming that the annular size of
the reprocessor is $R=100-10^5$$GM/c^2$, we find a 54.4--300 eV photon
flux of $\approx 4 \times 10^{46}$ photons s$^{-1}$, which matches the
flux that \citet{cbsk+08} find is required to power the ionization
cone\footnote{We obtained the value $4\times 10^{46}$ by integrating the
 photon flux, assuming blackbody emission, over the energy range from
 54.4 eV to 300 eV and over the entire radial extent of the
 reprocessing region.  We used the radial temperature profile specified
 by Equations (A3) and (A5) in Appendix A of \citet{osmt+08}.  Our
 result is insensitive to the values we assumed for the inner and outer
 radii of the annular reprocessor.  For example, if we halve the inner
 radius or double the outer radius, the 54.4 eV to 300 eV photon flux
 changes by only $\approx 10$\%.}.

What is the origin of the high spin of LMC X-1's black hole primary?
Was the black hole born with its present spin, or was it torqued up
gradually via the accretion flow supplied by its companion?  In order
to achieve a spin of $a_* \approx 0.9$ via disk accretion, an
initially non-spinning black hole must accrete more than $4 M_\odot$
from its donor \citep{kk+99} in becoming the $M = 10.9 M_\odot$ black
hole that we observe today \citep{osmt+08}.  However, to transfer this
much mass even in the case of Eddington-limited accretion
($\dot{M}_{Edd} \equiv L_{Edd}/c^2 \approx 2.50\times10^{-8} M_\odot/
{\rm yr}$) requires $> 150$ million years, whereas the age of the
system is only about 5 million years \citep{osmt+08}.  Thus, we
conclude that the spin of LMC X-1 is natal, which is the same
conclusion that has been reached for three other stellar black holes
\citep{msnr+06,smnd+06,lmnd+08}.  We note, however, that an
alternative scenario for the origin of high spins for black holes in
X-ray binaries has been proposed that invokes hypercritical accretion
\citep{mblp+08}.

\section{Conclusion}

We find that the spin of the black hole primary in LMC X-1 is $a_* =
0.92_{-0.07}^{+0.05}$, where the uncertainty includes all sources of
observational error and model parameter uncertainties.  This result is
based on an analysis of three thermal dominant (TD) spectra, which we
refer to as our gold spectra, plus 15 silver spectra.  The
observations that yielded the 15 silver spectra satisfy the TD
criterion that their rms power is $< 0.075$.  However, when analyzed
using the standard power-law model they fail to meet the TD-criterion
that their disk fractions are $> 75$\%.  We show that the failure of
the silver spectra to meet this criterion is caused by the unphysical
behavior of the standard power-law model, which diverges at low
energies and therefore intrudes on the thermal component of emission.
We sidestep this problem by using our Comptonization model {\sc
  simpl}.  If we admit an additional 35 bronze spectra (with rms power
$> 0.075$) into the analysis and consider all 53 {\it RXTE} spectra,
we obtain a consistent, somewhat higher, and less certain value of the
spin parameter. Meanwhile, our spin results for the 18 gold/silver
{\it RXTE} spectra are confirmed by our analyses of {\it XMM-Newton},
{\it BeppoSAX} and {\it Ginga} spectra (see Appendix A).

\acknowledgements

We thank an anonymous referee for very helpful comments and a thorough
reading of our paper. JEM acknowledges support from NASA grant
NNX08AJ55G. JFL acknowledges support provided by NASA through grant
HST-GO-11227 and the Chandra Fellowship Program, grant PF6-70043. RN
acknowledges support from NASA grant NNX08AH32G and NSF grant
AST-0805832. JFS was supported by the Smithsonian Endowment Funds. SWD
acknowledges support through the Chandra Fellowship Program, grant
PF6-70043. RAR acknowledges support from the NASA contract to MIT for
the instrument team contributing to {\it RXTE}. This research has made
use of data obtained from the High Energy Astrophysics Science Archive
Research Center (HEASARC) at NASA/Goddard Space Flight Center.

\appendix

\section{Determinations of the Spin Parameter for Five Additional X-ray
  Missions}
\label{other_mission_spin}

As stated in the final paragraph of \S5.1, our featured result for the
18 gold/silver {\it RXTE} spectra for the case $\alpha=0.01$, $a_* =
0.938 \pm 0.020$ (std.\ dev.; N=18), is confirmed by an analysis of
data from three other X-ray missions.  In the paragraphs below, which
are prefaced by a mission name, we consider in turn X-ray spectral
data for LMC X-1 that were obtained during the course of the five
missions named in Table \ref{other_mission} (column 2).  With the
exception of the BBXRT and {\it Ginga} data, all the data were
downloaded from the NASA HEASARC.  As discussed below, after applying
the data selection criterion developed in \S5.5, namely scattered
fraction $f_{\rm SC} < 4\%$, we find that only three of these missions
provide reliable estimates of spin (Table \ref{other_mission}, lines
1--3), i.e., estimates of spin that are comparable in quality to the
18 gold/silver {\it RXTE} PCA spectra.  As in the analysis of the {\it
  RXTE} data, for all five of these missions we fix the power-law
index at 2.5, which is a necessity for four of the five missions whose
detectors are unresponsive above $\sim 10$ keV.  As indicated below,
small corrections were applied to the effective areas of the various
detectors to align them with the standard \citet{ts+74} spectrum of
the Crab.  We consider here only the case $\alpha=0.01$. 

{\bf XMM-Newton:} A single 4940~s observation was performed on UT 2000
October 21 using the EPIC-pn camera in timing mode.  The data, which
are free of pile-up effects, were reduced using precisely the
procedures recommended for timing data
\footnote{ftp://ftp.xray.mpe.mpg.de/xmm/service/data\_analysis/EPIC\_PN/timing/timing\_mode.html}.
We fitted the data over the energy range 1.0--10.0 keV~\citep{wnps+03}
and corrected the detector's effective area by the factor 0.889 (i.e.,
the effective area is increased by dividing by the factor 0.889),
which is based on observations of the Crab \citep{kbb+05}.
Conclusion: The data meet the requirement $f_{\rm SC} < 4\%$, and the
spin is consistent with the {\it RXTE} value to within 0.9 standard
deviations (Table \ref{other_mission}, line 1).

{\bf BeppoSAX:} A single observation was made on UT 1997 October 5
using the narrow-field instruments aboard {\it BeppoSAX}.  The four detector
systems, as well as this specific observation of LMC X-1, are
described in detail in \citet{hgtc+01}.  We have chosen to analyze
only the data obtained by the Medium Energy Concentrator Spectrometer
(MECS = MECS2 + MECS3) because the PDS and HPGSPC spectra are noisy
and the LECS spectrum lacks a proper response file.  The MECS data
have been reduced following the standard procedures described in the
Cookbook for the {\it BeppoSAX} NFI Spectral
Analysis\footnote{http://heasarc.gsfc.nasa.gov/docs/sax/abc/saxabc/saxabc.html}.
The net eposure time was 40 ks.  The spectra obtained by the two MECS
detectors, MECS2 and MECS3, were fitted jointly over the energy range
1.65--10.5 keV; no correction to the effective area was required.
Conclusion: The data meet the requirement $f_{\rm SC} < 4\%$, and the
spin is consistent with the {\it RXTE} value to within 1.5 standard
deviations (Table \ref{other_mission}, line 2).

{\bf Ginga:} The {\it Ginga} Large Area Proportional Counter (LAC;
1.2-37 keV) observed LMC X-1 on six UT dates: 1987 April 22, 1987 July
16, 1987 September 30, 1988 August 15, 1990 June 21 and 1991 February
15.  The respective exposure times were approximately 7290, 12831,
4841, 315, 326, and 266 s.  We extracted the spectra following the
procedures described in \citet{emi+89}.  We applied to each spectrum
an effective-area correction factor of 0.968 based on an observation
of the Crab~\citep{ttpr+89}.  We fitted all six spectra; however, in
Table~\ref{other_mission} we report our results only for the spectrum
obtained on 1987 September 30, the only spectrum that satisfies our
data selection criterion ($f_{\rm SC}<4\%$).  This spectrum was fitted
over the restricted energy range 1.2--12.0 keV because the spectrum
contains large and unexplained residual features at higher energies.
Conclusion: The data meet the requirement $f_{\rm SC} < 4\%$, and the
spin is consistent with the {\it RXTE} value to within 0.8 standard
deviations (Table \ref{other_mission}, line 3).

{\bf BBXRT/Astro-1:} Four observations were performed on UT 1990
December 6 \& 7.  We disregard the observations on December 7 because
of their high background levels and large off-axis angles; we consider
only the pair of observations performed on December 6 with exposure
times of 1235~s and 616~s.  The data have been reduced following the
standard procedures \citep{smms+94}.  The two spectra were fitted
jointly over the energy range 0.7--7.0 keV, and the effective area was
corrected by the factor 0.973 \citep{wab+95}.  Conclusion: The
scattered fraction fails to meet the requirement $f_{\rm SC} < 4\%$,
the spin value is unreliable, and we disregard it (Table
\ref{other_mission}, line 4). However, we note that the BBXRT spin
value is consistent with the {\it RXTE} value to within 1.0 standard
deviation.

{\bf ASCA:} A single 3465~s observation was performed on UT 1995 April
2.  We consider only the data obtained by the Gas Imaging Spectrometer
(GIS).  We extracted the spectra following the standard procedures
described in the {\it ASCA} ABC guide
\footnote{http://heasarc.gsfc.nasa.gov/docs/asca/abc/abc.html}.
The two GIS spectra (GIS2 and GIS3) were fitted jointly over the
energy range 0.7--8.0 keV.  No correction to the effective area is
required because the GIS effective area calibrations were based on the
\citet{ts+74} spectrum of the Crab.  Conclusion: The scattered
fraction fails to meet the requirement $f_{\rm SC} < 4\%$, and the
spin value is unreliable.  We therefore disregard it while noting that
it is exceptionally low, a result for which we have no explanation.

\section{Contamination by the Pulsar PSR B0540--69}
\label{contamination_pulsar}

The pulsar PSR B0540--69 is located 25$'$ from LMC X-1.
\citet{hgtc+01} concluded that their {\it BeppoSAX} PDS spectra
(15--100 keV) of LMC X-1 were contaminated by the presence of this
nearby pulsar.  Even though the PCA has a narrower field-of-view than
the {\it BeppoSAX} PDS, 1\fdg0 versus 1\fdg5 (FWHM), our spectra of
LMC X-1 could likewise be significantly contaminated.  We now describe
three lines of evidence showing that this is not the case: We first
estimate the contribution of the pulsar to the total flux using its
well-determined spectrum; we then search for the expected signature of
the pulsar's hard spectrum in our data; and lastly we subtract a faked
version of the pulsar's spectrum from the source spectrum.  (We
disregard PSR J0537--6910, which also lies in the PCA field of view,
because it is further off-axis and significantly fainter than PSR
B0540--69; \citealp{cimm+98}).

PSR B0540--69 and its $2''-3''$ plerion nebula were discovered 25
years ago by the {\it Einstein} X-ray observatory \citep{shh+84}.  The
source has since been closely observed by many missions (e.g., {\it
  ROSAT}, {\it RXTE}, {\it Chandra}, and {\it INTEGRAL}).  It is quite
similar to the Crab.  Like the Crab, all studies to date show that its
hard spectrum can be fitted with a single power law with a photon
index $\Gamma$ in the range 1.8 to 2.2
\citep{sgcl+90,kmag+01,gmms+06}.  On the other hand, the power-law
component in the spectrum of LMC X-1, is much softer, $\Gamma \sim
2.5-3$ \citep[Table 1;][]{nwhp+01,hgtc+01}.  We now estimate at what
energy the flux in the flat pulsar spectrum first exceeds the flux of
LMC X-1.  With LMC X-1 centered in the PCA, the collimator
transmission factor for the pulsar is $T = 0.57$.  Adopting for the
pulsar $\Gamma=2$ and a flux at 1 keV of $A = 8.86 \times 10^{-3}$ ph
cm$^{-2}$~s$^{-1}$~keV$^{-1}$ (based on the flux observed by {\it
  INTEGRAL} in 2003; \citealp{gmms+06}), the expected pulsar photon
flux is $8.9 \times 10^{-3} \times T \times
E^{-2}=5.05\times10^{-3}E^{-2}$.  For LMC X-1 we adopt $\Gamma=2.5$
and $A=0.2$ (a typical value in Table~1), for which the corresponding
photon flux is $0.2 \times E^{-2.5}$.  Equating the two fluxes, the
flux of the pulsar dominates over that of LMC X-1 only at energies
$\gtrsim 1$~MeV; even for $\Gamma$(LMC~X-1)$=3$, the pulsar dominates
only for energies $\gtrsim$ 40 keV.

We now describe a second test performed on our 18 gold/silver spectra
which further demonstrates that the flux from the nearby pulsar is
unimportant.  In addition to fitting the spectra over our adopted energy
range of 2.5--20.0 keV (\S\S3,4), we also fitted them over the ranges
2.5--15.0 keV and 2.5--25.0 keV using precisely the same model and
procedures.  The results are displayed in Figure
\ref{pulsar_effect} where it can be seen that the fit parameters
and the values of $\chi_{\nu}^{2}$ are scarcely affected by the choice
of the energy range.  This is not at all what one would expect if the
spectrum of LMC X-1 was contaminated by a significant component of flux
from the pulsar.  Rather, one would expect the fit results to vary as
the upper energy bound is increased from 15 to 25 keV because the
pulsar's spectrum is significantly harder than that of LMC X-1 (see
above).  Specifically, one would expect the influence of the pulsar to
be most apparent at 25 keV.  However, the effect of the pulsar on our
results, if any, are seen to be negligible.  In particular, the values
of $a_*$ (top panel) are completely insensitive to the choice of the
fitting range.

As a final direct test, we created a ``faked'' spectrum of the pulsar
in XSPEC using the observed flux and power-law index given above and
including the off-axis transmission factor ($T = 0.57$).  We then
subtracted this simulated spectrum from each of the 18 gold/silver
spectra, redid our standard analysis and found $a_* = 0.947\pm0.020$,
which differs by only 0.3 standard deviations from our adopted result
($a_* = 0.938\pm0.020$; \S\ref{primary_results}), which was obtained
by ignoring the flux of the pulsar.

We thus conclude, based on the three arguments given above, that the
contaminating effect of the pulsar on our estimate of the spin
parameter of LMC X-1 is negligible ($\lesssim 1$\%).


\newpage

\begin{deluxetable}{cccccccccccccc}
\tabletypesize{\scriptsize} \rotate \tablewidth{0pt}
\tablecaption{Preliminary Results for LMC X-1 {\it RXTE} Data using {\sc
phabs(diskbb+powerlaw)\tablenotemark{a}}} \tablehead{ \colhead{Obs.\tablenotemark{b}} &
\colhead{UT} & \colhead{MJD\tablenotemark{c}} & \colhead{ $T_{\rm
obs}$} & \colhead{Count Rate}& \colhead{rms} & \colhead{$T_{\rm in}$}
& \colhead{K\tablenotemark{e}} & \colhead{$\Gamma$\tablenotemark{f}} &
\colhead{Power-Law} & \colhead{DF\tablenotemark{g}} &
\colhead{State\tablenotemark{h}} & \colhead{$\chi^2_{\nu}$/dof} \cr
\colhead{No.}  & \colhead{(yyyy-mm-dd)} & \colhead{} & \colhead{(s)} &
\colhead{($\rm cts/s$)} & \colhead{($10^{-2})$\tablenotemark{d}} &
\colhead{(keV)} & \colhead{} & \colhead{} & \colhead{Norm} &
\colhead{(\%)} & \colhead{} & \colhead{} \cr \colhead{(1)} &
\colhead{(2)} & \colhead{(3)} & \colhead{(4)} & \colhead{(5)} &
\colhead{(6)} & \colhead{(7)} & \colhead{(8)} & \colhead{(9)} &
\colhead{(10)} & \colhead{(11)} & \colhead{(12)} & \colhead{(13)} }
\startdata \hline \hline 1 & 1996-06-09& 50243.888 & 4960&$ 31.27\pm
0.16$&$ 7.43\pm0.64$& $0.864\pm0.008$&$52.73\pm 4.14$&$ 3.06\pm 0.14$&
$0.20\pm 0.06$&64.45&SPL:TD& 0.93/43 \\ 2 & 1996-08-01& 50296.917 &
4960&$ 34.71\pm 0.17$&$ 7.24\pm0.77$& $0.846\pm0.007$&$74.05\pm
3.91$&$ 2.91\pm 0.14$& $0.15\pm 0.05$&70.92&SPL:TD& 0.62/43 \\ 3 &
1996-12-06& 50423.854 & 73168&$ 33.20\pm 0.10$&$ 6.17\pm0.17$&
$0.918\pm0.004$&$45.11\pm 1.42$&$ 3.13\pm 0.07$& $0.20\pm
0.03$&70.83&SPL:TD& 1.28/43 \\ 4 & 1997-03-09& 50516.454 & 10032&$
33.44\pm 0.14$&$ 6.47\pm0.44$& $0.911\pm0.007$&$45.75\pm 3.10$&$
3.17\pm 0.12$& $0.23\pm 0.06$&67.90&SPL:TD& 1.05/43 \\ 5 & 1997-03-21&
50528.108 & 10432&$ 34.25\pm 0.14$&$ 6.33\pm0.43$&
$0.914\pm0.007$&$47.17\pm 3.29$&$ 3.28\pm 0.12$& $0.26\pm
0.07$&69.49&SPL:TD& 0.74/43 \\ 6 & 1997-04-16& 50554.291 & 11312&$
34.48\pm 0.13$&$ 6.05\pm0.46$& $0.889\pm0.007$&$51.97\pm 3.57$&$
3.28\pm 0.10$& $0.30\pm 0.07$&65.42&SPL:TD& 0.89/43 \\ 7 & 1997-05-07&
50575.107 & 9744&$ 34.91\pm 0.14$&$ 7.15\pm0.42$&
$0.906\pm0.006$&$49.90\pm 3.22$&$ 3.19\pm 0.12$& $0.24\pm
0.06$&69.06&SPL:TD& 0.72/43 \\ 8 & 1997-05-28& 50596.944 & 4400&$
32.26\pm 0.17$&$ 6.09\pm0.63$& $0.907\pm0.011$&$41.24\pm 4.56$&$
3.25\pm 0.13$& $0.29\pm 0.09$&62.38&SPL:TD& 0.66/43 \\ 9 & 1997-05-29&
50597.047 & 2768&$ 30.55\pm 0.20$&$ 7.48\pm0.88$&
$0.884\pm0.011$&$51.28\pm 5.70$&$ 3.19\pm 0.21$& $0.20\pm
0.10$&70.71&SPL:TD& 0.81/43 \\ 10 & 1997-07-09& 50638.653 & 4240&$
31.12\pm 0.17$&$ 5.74\pm0.76$& $0.860\pm0.009$&$59.68\pm 6.12$&$
3.28\pm 0.20$& $0.26\pm 0.12$&67.89&SPL:TD& 0.70/42 \\ 11 &
1997-08-20& 50680.829 & 9056&$ 36.30\pm 0.15$&$ 7.04\pm0.42$&
$0.922\pm0.007$&$47.88\pm 3.55$&$ 3.26\pm 0.14$& $0.28\pm
0.09$&68.82&SPL:TD& 0.79/42 \\ 12 & 1997-09-12& 50703.106 & 9456&$
34.66\pm 0.14$&$ 5.61\pm0.53$& $0.876\pm0.007$&$68.10\pm 2.62$&$
2.85\pm 0.17$& $0.10\pm 0.04$&78.75&TD& 0.68/42 \\ 13 & 1997-09-19&
50710.736 & 9680&$ 33.55\pm 0.14$&$ 7.34\pm0.54$&
$0.920\pm0.007$&$43.29\pm 2.80$&$ 3.09\pm 0.13$& $0.21\pm
0.07$&67.41&SPL:TD& 0.71/42 \\ 14 & 1997-12-12& 50794.380 & 9616&$
34.14\pm 0.14$&$ 5.73\pm0.49$& $0.893\pm0.007$&$57.40\pm 3.88$&$
3.27\pm 0.17$& $0.24\pm 0.09$&72.74&SPL:TD& 0.52/42 \\ 15 &
1998-03-12& 50884.606 & 9760&$ 36.46\pm 0.15$&$ 6.06\pm0.4$&
$0.949\pm0.008$&$39.62\pm 3.35$&$ 3.28\pm 0.13$& $0.32\pm
0.10$&65.78&SPL:TD& 0.84/42 \\ 16 & 1998-05-06& 50939.252 & 11248&$
32.54\pm 0.13$&$ 5.62\pm0.43$& $0.903\pm0.006$&$49.67\pm 3.13$&$
3.19\pm 0.15$& $0.21\pm 0.07$&70.92&SPL:TD& 0.94/42 \\ 17 &
1998-07-20& 51014.014 & 9936&$ 32.41\pm 0.13$&$ 5.97\pm0.51$&
$0.894\pm0.006$&$56.54\pm 2.82$&$ 3.04\pm 0.18$& $0.14\pm
0.06$&77.19&TD& 0.62/42 \\ 18 & 2004-01-07& 53011.559 & 9856&$
28.78\pm 0.13$&$ 6.61\pm0.88$& $0.892\pm0.008$&$57.53\pm 2.12$&$
2.42\pm 0.20$& $0.03\pm 0.01$&85.65&TD& 0.80/37 \\ \hline \hline 19 &
1996-04-18& 50191.325 & 4400&$ 39.07\pm 0.18$&$ 8.29\pm0.86$&
$0.848\pm0.009$&$59.05\pm 4.57$&$ 2.98\pm 0.09$& $0.30\pm
0.06$&52.52&SPL:TD& 0.91/43 \\ 20 & 1996-05-18& 50221.352 & 4896&$
36.22\pm 0.17$&$ 7.91\pm0.88$& $0.841\pm0.010$&$53.60\pm 4.65$&$
3.01\pm 0.09$& $0.32\pm 0.06$&48.77&SPL& 0.79/43 \\ 21 & 1996-07-05&
50269.123 & 4576&$ 33.57\pm 0.17$&$ 10.31\pm0.73$&
$0.997\pm0.014$&$21.78\pm 2.57$&$ 3.03\pm 0.10$& $0.26\pm
0.06$&53.73&SPL:TD& 1.07/43 \\ 22 & 1996-09-06& 50332.946 & 4912&$
33.78\pm 0.16$&$ 11.42\pm0.49$& $0.994\pm0.012$&$20.46\pm 2.07$&$
2.81\pm 0.09$& $0.20\pm 0.04$&50.03&SPL:TD& 1.42/43 \\ 23 & 1996-10-04&
50360.555 & 4832&$ 33.08\pm 0.16$&$ 8.04\pm0.58$&
$0.898\pm0.009$&$44.05\pm 3.90$&$ 3.07\pm 0.12$& $0.23\pm
0.07$&61.63&SPL:TD& 0.95/43 \\ 24 & 1996-12-30& 50447.533 & 9760&$
35.52\pm 0.14$&$ 9.39\pm0.41$& $0.982\pm0.010$&$26.09\pm 2.23$&$
3.11\pm 0.08$& $0.31\pm 0.05$&55.07&SPL:TD& 1.06/43 \\ 25 &
1997-01-18& 50466.436 & 9808&$ 39.21\pm 0.15$&$ 9.67\pm0.31$&
$0.998\pm0.010$&$25.28\pm 2.13$&$ 3.09\pm 0.07$& $0.34\pm
0.06$&53.18&SPL:TD& 0.99/43 \\ 26 & 1997-02-08& 50487.468 & 9664&$
39.49\pm 0.15$&$ 7.58\pm0.4$& $0.894\pm0.006$&$57.30\pm 2.92$&$
3.01\pm 0.10$& $0.23\pm 0.05$&66.06&SPL:TD& 1.40/43 \\ 27 &
1997-06-18& 50617.157 & 5120&$ 34.28\pm 0.17$&$ 7.86\pm0.65$&
$0.873\pm0.007$&$59.80\pm 4.50$&$ 3.14\pm 0.14$& $0.23\pm
0.07$&68.14&SPL:TD& 0.74/43 \\ 28 & 1997-07-04& 50633.313 & 2464&$
31.87\pm 0.21$&$ 8.42\pm0.96$& $0.846\pm0.009$&$72.21\pm 4.21$&$
2.73\pm 0.22$& $0.09\pm 0.05$&75.25&SPL:TD& 0.76/43 \\ 29 &
1997-08-01& 50661.887 & 8992&$ 33.14\pm 0.14$&$ 7.71\pm0.42$&
$0.948\pm0.009$&$32.62\pm 2.75$&$ 3.14\pm 0.12$& $0.27\pm
0.07$&60.74&SPL:TD& 0.99/42 \\ 30 & 1997-09-09& 50700.812 & 9968&$
34.62\pm 0.14$&$ 7.92\pm0.39$& $0.960\pm0.008$&$31.18\pm 2.41$&$
3.08\pm 0.10$& $0.26\pm 0.06$&59.73&SPL:TD& 1.24/42 \\ 31 &
1997-10-10& 50731.708 & 11264&$ 33.25\pm 0.13$&$ 9\pm0.35$&
$1.012\pm0.010$&$21.65\pm 1.96$&$ 3.12\pm 0.10$& $0.28\pm
0.06$&57.15&SPL:TD& 0.90/42 \\ 32 & 1997-11-01& 50753.728 & 9008&$
33.66\pm 0.14$&$ 7.63\pm0.41$& $0.989\pm0.008$&$29.25\pm 1.72$&$
2.86\pm 0.13$& $0.14\pm 0.04$&68.74&SPL:TD& 0.98/42 \\ 33 &
1997-11-23& 50775.661 & 8336&$ 35.17\pm 0.14$&$ 8.43\pm0.4$&
$0.975\pm0.009$&$30.07\pm 2.38$&$ 3.05\pm 0.12$& $0.24\pm
0.07$&61.82&SPL:TD& 0.99/42 \\ 34 & 1998-01-04& 50817.465 & 9344&$
40.35\pm 0.15$&$ 7.93\pm0.33$& $0.971\pm0.007$&$38.55\pm 2.61$&$
3.12\pm 0.12$& $0.27\pm 0.07$&66.60&SPL:TD& 1.27/42 \\ 35 &
1998-01-25& 50838.750 & 9856&$ 36.03\pm 0.14$&$ 7.96\pm0.32$&
$0.988\pm0.009$&$30.17\pm 2.48$&$ 3.17\pm 0.12$& $0.28\pm
0.08$&63.57&SPL:TD& 1.01/42 \\ 36 & 1998-02-20& 50864.466 & 9728&$
37.46\pm 0.14$&$ 9.44\pm0.35$& $0.994\pm0.009$&$26.72\pm 2.34$&$
3.12\pm 0.09$& $0.33\pm 0.07$&56.17&SPL:TD& 1.23/42 \\ 37 &
1998-04-07& 50910.312 & 9584&$ 37.91\pm 0.15$&$ 8.06\pm0.36$&
$0.952\pm0.009$&$34.39\pm 3.21$&$ 3.28\pm 0.09$& $0.42\pm
0.09$&55.77&SPL:TD& 0.86/42 \\ 38 & 1998-05-28& 50961.227 & 9872&$
38.37\pm 0.15$&$ 7.91\pm0.36$& $0.923\pm0.009$&$42.00\pm 3.69$&$
3.31\pm 0.10$& $0.44\pm 0.10$&56.82&SPL:TD& 0.67/42 \\ 39 &
1998-06-28& 50992.190 & 9760&$ 38.25\pm 0.15$&$ 7.99\pm0.34$&
$0.966\pm0.012$&$29.39\pm 3.42$&$ 3.42\pm 0.08$& $0.59\pm
0.11$&50.00&SPL:TD& 0.89/42 \\ 40 & 1998-08-13& 51038.848 & 9200&$
38.38\pm 0.15$&$ 8.92\pm0.35$& $0.971\pm0.010$&$29.09\pm 2.89$&$
3.21\pm 0.09$& $0.43\pm 0.09$&51.89&SPL:TD& 0.82/42 \\ 41 &
1998-09-02& 51058.798 & 9264&$ 35.71\pm 0.15$&$ 7.87\pm0.4$&
$0.940\pm0.011$&$34.45\pm 3.70$&$ 3.35\pm 0.10$& $0.45\pm
0.10$&55.07&SPL:TD& 1.01/42 \\ 42 & 1998-09-29& 51085.826 & 7120&$
33.14\pm 0.14$&$ 9.33\pm0.59$& $0.964\pm0.013$&$25.16\pm 3.21$&$
3.25\pm 0.10$& $0.41\pm 0.09$&49.87&SPL& 1.16/42 \\ 43 & 1998-09-30&
51086.525 & 2992&$ 33.84\pm 0.20$&$ 10.89\pm0.57$&
$1.089\pm0.030$&$11.02\pm 2.44$&$ 3.12\pm 0.10$& $0.40\pm
0.09$&40.62&SPL& 0.98/42 \\ 44 & 2004-01-08& 53012.266 & 8688&$
31.24\pm 0.13$&$ 8.66\pm0.58$& $0.960\pm0.009$&$30.75\pm 2.40$&$
3.06\pm 0.12$& $0.22\pm 0.06$&62.82&SPL:TD& 1.00/37 \\ 45 &
2004-01-08& 53012.749 & 5360&$ 30.34\pm 0.15$&$ 7.83\pm0.66$&
$0.973\pm0.011$&$28.62\pm 3.04$&$ 3.16\pm 0.16$& $0.23\pm
0.09$&65.01&SPL:TD& 0.66/37 \\ 46 & 2004-01-09& 53013.166 & 9264&$
31.20\pm 0.13$&$ 8.9\pm0.54$& $0.977\pm0.010$&$25.97\pm 2.45$&$
3.17\pm 0.11$& $0.29\pm 0.07$&58.06&SPL:TD& 1.05/37 \\ 47 &
2004-01-09& 53013.801 & 3168&$ 30.84\pm 0.18$&$ 8.19\pm0.95$&
$0.901\pm0.009$&$51.84\pm 4.36$&$ 3.05\pm 0.26$& $0.14\pm
0.09$&75.47&SPL:TD& 0.72/37 \\ 48 & 2004-01-10& 53014.184 & 10160&$
30.97\pm 0.13$&$ 8.61\pm0.65$& $0.948\pm0.008$&$33.78\pm 2.51$&$
3.12\pm 0.12$& $0.23\pm 0.07$&64.73&SPL:TD& 1.15/37 \\ 49 &
2004-01-10& 53014.783 & 6480&$ 32.81\pm 0.15$&$ 9.2\pm0.59$&
$0.975\pm0.011$&$27.61\pm 2.88$&$ 3.17\pm 0.12$& $0.30\pm
0.08$&58.16&SPL:TD& 0.59/37 \\ 50 & 2004-01-11& 53015.428 & 10800&$
31.06\pm 0.13$&$ 7.75\pm0.56$& $0.946\pm0.010$&$32.07\pm 3.10$&$
3.33\pm 0.11$& $0.36\pm 0.09$&59.00&SPL:TD& 0.82/37 \\ 51 &
2004-01-11& 53015.871 & 4064&$ 33.19\pm 0.17$&$ 11\pm0.61$&
$1.046\pm0.013$&$18.82\pm 2.00$&$ 2.83\pm 0.13$& $0.18\pm
0.06$&57.70&SPL:TD& 0.74/37 \\ 52 & 2004-01-12& 53016.340 & 10704&$
32.75\pm 0.13$&$ 10.23\pm0.44$& $1.051\pm0.010$&$17.18\pm 1.45$&$
2.91\pm 0.09$& $0.22\pm 0.05$&54.11&SPL:TD& 0.85/37 \\ 53 &
2004-01-12& 53016.822 & 4720&$ 31.45\pm 0.17$&$ 11.74\pm0.59$&
$1.170\pm0.029$&$8.28\pm 1.47$&$ 3.22\pm 0.09$& $0.42\pm
0.08$&44.71&SPL& 0.70/37 \\ \hline \enddata 
\tablenotetext{a}{Column density fixed at $N_{\rm H} = 4.6 \times
10^{21}$ cm$^{-2}$.}
\tablenotetext{b}{The
first 18 spectra, which meet the TD criterion rms $<$ 0.075, comprise
our favored sample of gold/silver spectra.}
\tablenotetext{c}{Midpoint time of observation.  MJD = JD -
2,400,000.5.}  \tablenotetext{d}{Total rms power integrated over
0.1--10 Hz in the power density spectrum (2--30 keV).}
\tablenotetext{e}{Normalization constant of the thermal component: $K
\propto R_{\rm in}^2$, where $R_{\rm in}$ is the inner-disk radius.}
\tablenotetext{f}{Photon index of the power-law component.}
\tablenotetext{g}{Disk fraction defined over the energy range 2--20 keV.}
\tablenotetext{h}{See \S1 and text in \citet{rm+06} for state
definitions.}
\end{deluxetable}

\begin{table}[ht]
\scriptsize
\caption{Primary Results for LMC X-1 {\it RXTE} Data using {\sc
    phabs(simpl$\otimes$kerrbb{\small 2})}\tablenotemark{a} }
\centering
\begin{tabular}{l c c c c c c c c c c c}
  \hline\hline No.&MJD &a* &$ \dot{M} $\tablenotemark{b}& f &$ \Gamma $&
  $f_{\rm{SC}}$ &NDF(\%) &$ \chi^2_{\nu}$/dof\tablenotemark{c} &$
  L/L_{edd} $ \\ \hline 
  1& 50243.888 &$0.933\pm0.007$ &$
  1.256\pm0.040$&1.557&$2.5$&$0.049\pm0.002$&85.87&0.89/44 &0.145 \\
  \hline 2& 50296.917 &$0.884\pm0.008$ &$
  1.679\pm0.047$&1.551&$2.5$&$0.044\pm0.001$&86.84&0.56/44 &0.164 \\
  \hline 3& 50423.854 &$0.962\pm0.003$ &$
  1.183\pm0.021$&1.567&$2.5$&$0.035\pm0.001$&90.29&1.37/44 &0.158 \\
  \hline 4& 50516.454 &$0.949\pm0.005$ &$
  1.262\pm0.032$&1.573&$2.5$&$0.038\pm0.001$&89.36&1.04/44 &0.157 \\
  \hline 5& 50528.108 &$0.944\pm0.005$ &$
  1.333\pm0.031$&1.574&$2.5$&$0.032\pm0.001$&90.93&0.80/44 &0.162 \\
  \hline 6& 50554.291 &$0.928\pm0.005$ &$
  1.431\pm0.032$&1.566&$2.5$&$0.037\pm0.001$&89.35&0.98/44 &0.161 \\
  \hline 7& 50575.107 &$0.938\pm0.004$ &$
  1.379\pm0.031$&1.573&$2.5$&$0.036\pm0.001$&89.84&0.74/44 &0.162 \\
  \hline 8& 50596.944 &$0.952\pm0.006$ &$
  1.193\pm0.040$&1.569&$2.5$&$0.043\pm0.002$&87.98&0.72/44 &0.151 \\
  \hline 9& 50597.047 &$0.939\pm0.007$ &$
  1.260\pm0.045$&1.563&$2.5$&$0.033\pm0.002$&90.26&0.78/44 &0.150 \\
  \hline 10& 50638.653 &$0.916\pm0.009$ &$
  1.437\pm0.054$&1.550&$2.5$&$0.033\pm0.002$&89.98&0.66/43 &0.155 \\
  \hline 11& 50680.829 &$0.945\pm0.005$ &$
  1.401\pm0.037$&1.572&$2.5$&$0.033\pm0.001$&90.83&0.75/43 &0.170 \\
  \hline 12\tablenotemark{d}& 50703.106 &$0.909\pm0.007$ &$
  1.594\pm0.043$&1.551&$2.5$&$0.032\pm0.001$&90.44&0.58/43 &0.168 \\
  \hline 13& 50710.736 &$0.966\pm0.005$ &$
  1.160\pm0.035$&1.554&$2.5$&$0.042\pm0.001$&88.54&0.67/43 &0.159 \\
  \hline 14& 50794.380 &$0.926\pm0.006$ &$
  1.499\pm0.039$&1.561&$2.5$&$0.028\pm0.001$&91.79&0.57/43 &0.168 \\
  \hline 15& 50884.606 &$0.965\pm0.004$ &$
  1.257\pm0.036$&1.569&$2.5$&$0.036\pm0.001$&90.37&0.83/43 &0.171 \\
  \hline 16& 50939.252 &$0.947\pm0.005$ &$
  1.295\pm0.036$&1.562&$2.5$&$0.033\pm0.001$&90.56&0.84/43 &0.160 \\
  \hline 17\tablenotemark{d}& 51014.014 &$0.936\pm0.005$ &$
  1.374\pm0.037$&1.561&$2.5$&$0.029\pm0.001$&91.50&0.49/43 &0.161 \\
  \hline 18\tablenotemark{d}& 53011.559 &$0.940\pm0.005$ &$
  1.273\pm0.036$&1.551&$2.5$&$0.030\pm0.001$&91.03&0.76/38 &0.151 \\
  \hline \hline 19&&$0.938\pm0.020$\tablenotemark{e} &$
  1.348\pm0.137$&$1.562\pm0.008$&$2.5$&$0.036\pm0.005$&$89.76\pm1.53$&$0.78\pm0.20
  $&$0.160\pm 0.007$ \\ \hline
\end{tabular}
\tablenotetext{a}{Column density fixed at $N_{\rm H} = 4.6 \times
  10^{21}$ cm$^{-2}$ and with viscosity parameter $\alpha=0.01$.}
\tablenotetext{b}{The mass accretion rate is in units of
  $10^{18}$~g~$\rm s^{-1}$.}  \tablenotetext{c}{ The number of degrees of freedom
  (dof) in Table 1 is one less because here $\Gamma$ is fixed at 2.5.}
\tablenotetext{d}{The three strictly TD-state spectra; their average
  spin parameter is $a_*=0.928\pm0.014$.}  \tablenotetext{e}{The value
  of the spin parameter was computed using the one-sided Green's
  function {\sc simpl-1} (\S4.2).  The two-sided Green's function gives
  an essentially identical result with a mean spin parameter
  $a_*=0.937\pm0.020$.  If one includes a Gaussian emission line in the
  fit to all 18 spectra, {\sc phabs(simpl$\otimes$(kerrbb2+gaussian))},
  the average spin parameter decreases slightly by 0.6
  standard deviations to $a_* = 0.922\pm0.021$ (\S5.2).}
\label{gold_spectra_result}
\end{table}

\begin{table}[ht]
\caption{Comparison of Fit Results for Different Models and Data
Samples} \centering \scriptsize
\begin{tabular}{c c c c c c c}
\hline\hline
 Model& $a_*$ &$ \dot{M} $& $ \Gamma $& $f_{\rm SC}$ &NDF(\%) &Comments \\ 
\hline
{\sc phabs(simpl}$\otimes${\sc kerrbb2)}\tablenotemark{a} & $0.938\pm0.020$ & $1.348\pm0.137$ & 2.5&
 $0.036\pm0.005$& $89.76\pm1.53$& 18 spectra with $\Gamma$ fixed \\
\hline
\hline
{\sc phabs(simpl}$\otimes${\sc kerrbb2)} &$0.929\pm0.020$ & $1.388\pm0.136$&
 $2.76\pm0.29$& $0.050\pm0.016$&  $88.55\pm2.16$& 18 spectra with $\Gamma$ free \\
\hline
{\sc phabs(kerrbb2+powerlaw)} & $0.953\pm0.018$ & $1.193\pm0.133$ & 2.5&
-- & $84.11\pm2.28$& 18 spectra with $\Gamma$ fixed \\
\hline
{\sc phabs(simpl}$\otimes${\sc kerrbb2)} &$0.961\pm0.033$ & $1.137\pm0.293$& 2.5&
 $0.051\pm0.018$&  $86.82\pm3.62$& 53 spectra with $\Gamma$ fixed \\
\hline
\end{tabular}
\tablenotetext{a}{$a_*=0.944\pm0.019$ for $\Gamma$ fixed at 2.3
  and $a_*=0.932\pm0.021$ for $\Gamma$ fixed at 2.7. In the extreme
  case of $\Gamma=3.1$, $a_*=0.918\pm0.024$.}
\label{rxte_final_results}
\end{table}

\begin{table}[ht]
\scriptsize
\caption{Results for LMC X-1 for Five Other X-ray Missions using {\sc
    phabs(simpl$\otimes$kerrbb{\small 2})}\tablenotemark{a}}
\centering
\begin{tabular}{c c c c c c c c c c c c }
  \hline\hline No.&Mission \& Detector &a* &$ \dot{M} $& f &$ \Gamma $& $f_{\rm
    SC}$ &NDF(\%)&$ \chi^2_{\nu} $&$ L/L_{edd} $ &$N_{\rm H}$(1.0E22)  \\ \hline
  1& {\it XMM-Newton} EPIC &$0.957\pm0.002$
  &$
  0.746\pm0.011$&1.618&$2.5$&$0.002\pm0.002$&99.35&0.96
  &0.123 & $0.443\pm0.004$  \\ \hline 2& {\it BeppoSAX} MECS
  &$0.969\pm0.006$ &$
  1.057\pm0.039$&1.628&$2.5$&$0.021\pm0.003$&94.18&1.03
  &0.148 & $0.459\pm0.031$  \\ \hline
  3
  &{\it Ginga} LAC\tablenotemark{a}&$0.914\pm0.022$&$1.487\pm0.136$&1.606&2.5&$0.023\pm0.005$&93.20&0.97&0.159&$0.453\pm0.093$ \\
  \hline \hline 4& BBXRT/Astro-1
  &$0.902\pm0.028$ &$
  0.934\pm0.067$&1.605&$2.5$&$0.099\pm0.019$&71.90&1.06
  &0.096 & $0.575\pm0.015$  \\ \hline 5& {\it ASCA} GIS\tablenotemark{b}
  &$0.778\pm0.012$ &$
  2.287\pm0.048$&1.643&$2.5$&$0.079\pm0.005$&78.64&0.70
  &0.176 & $0.512\pm0.006$  \\
  \hline
\end{tabular}
\label{other_mission}
\tablenotetext{a}{In order to fit this {\it Ginga} LAC spectrum, it was 
  necessary to add a Gaussian emission-line component, {\sc
    phabs(simpl$\otimes$(kerrbb{\small 2} + gaussian))}; the central
  energy of the line, its full-width at half maximum and equivalent
  width are respectively  $7.0
  \pm 1.5$~keV, $1.3 \pm 0.6$~keV and 0.51~keV.}\tablenotetext{b}{{\it ASCA}
  SIS gives a similarly low value of spin $a_*=0.742\pm0.012$.}
\end{table}

\clearpage

\begin{figure}[ht]
\centering
\plotone{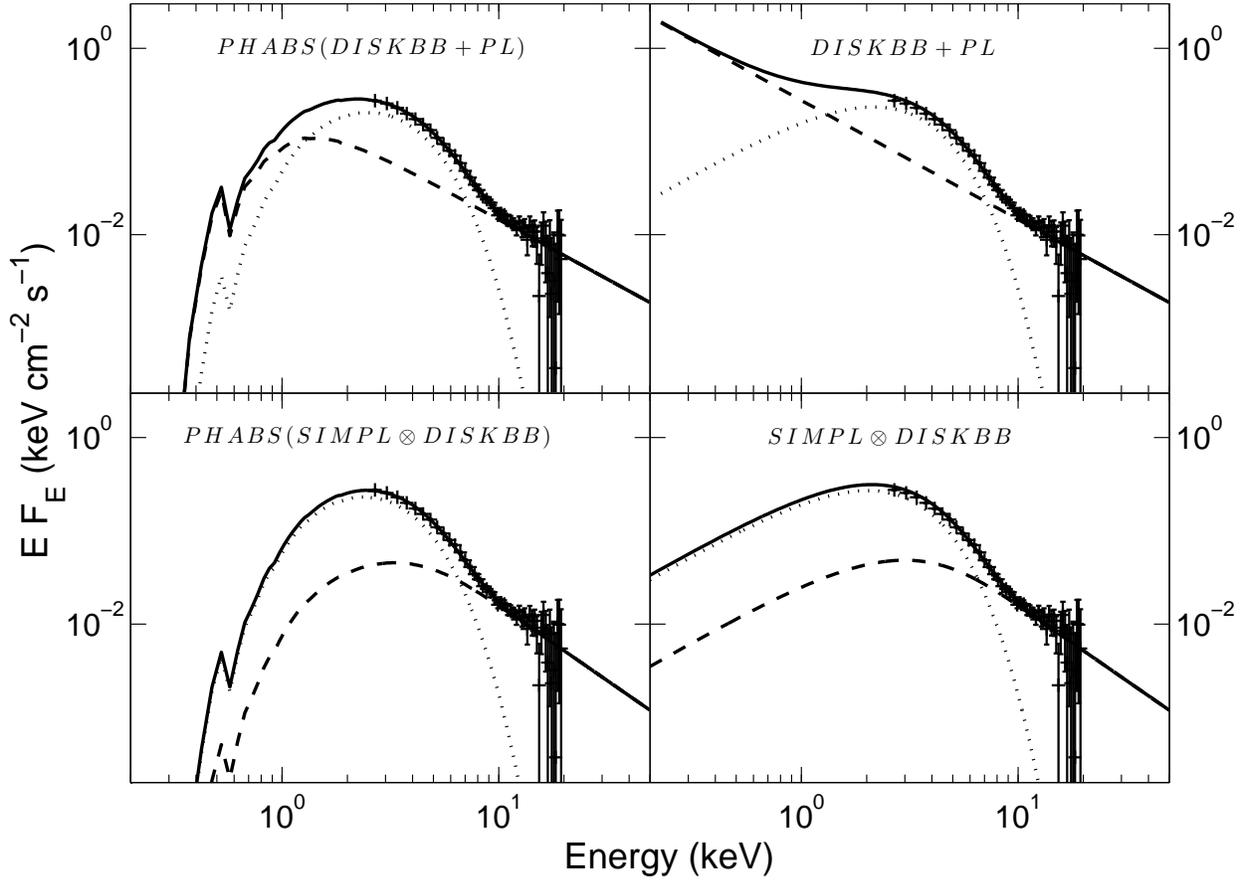}
\caption{{\sc powerlaw} versus {\sc simpl}.  The left pair of panels
show unfolded spectral fits to a PCA observation of LMC X-1 (No.\ 5 in
Table 1) and the pair on the right show the corresponding unabsorbed
models.  All of the fits were performed using {\sc diskbb} in
conjunction with {\sc powerlaw} (top panels) and {\sc simpl} (lower
panels).  The composite model is represented by a solid line, the
emergent disk component by a dotted line, and the Comptonization
component by a dashed line.  Note the strikingly different behaviors
of this latter component in the unabsorbed models.}
\label{simpl_pl_difference}
\end{figure}

\clearpage

\begin{figure}[ht]
\centering
\plotone{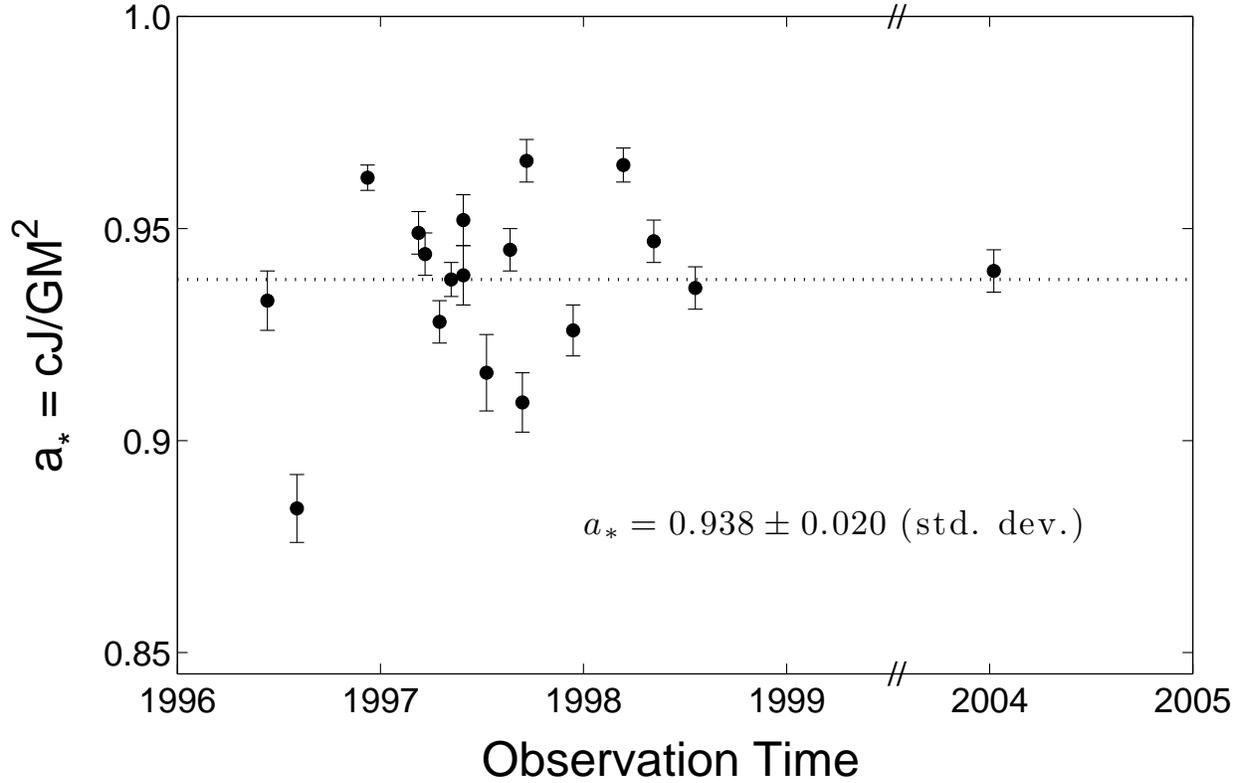}
\caption{Spin parameter versus observing time.  The dashed line
  indicates the mean value of the spin parameter for the 18 gold/silver
  spectra.  As indicated, the scatter about the mean -- an effective
  measure of the observational uncertainty -- is $\approx 2$\%.  The
  small error bars shown reflect only the counting-statistical errors
  associated with the many counts detected per observation ($\gtrsim 10^5$
  counts; Table 1).}
\label{evolution_spin_parameter}
\end{figure}

\newpage

\begin{figure}[ht]
\centering
\plotone{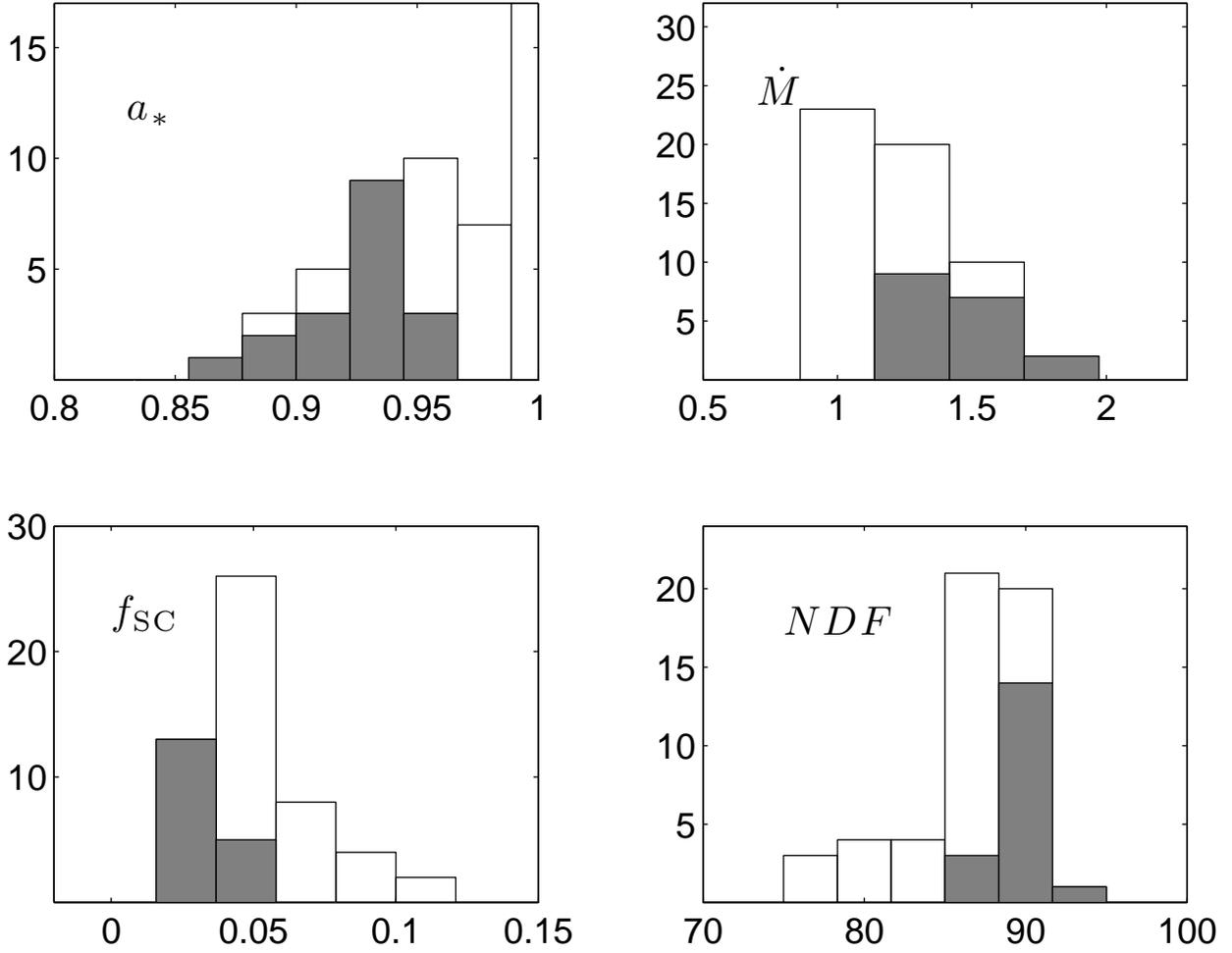}
\caption{Histograms of the fitting parameters for our adopted model
{\sc phabs(simpl}$\otimes${\sc kerrbb{\small 2}}) with $\Gamma$ fixed
at 2.5.  The solid histograms are for the 18 gold/silver spectra and the
open plus solid histograms are for all 53 gold/silver/bronze spectra.}
\label{rxte_histogram}
\end{figure}

\begin{figure}[ht]
\centering
\plotone{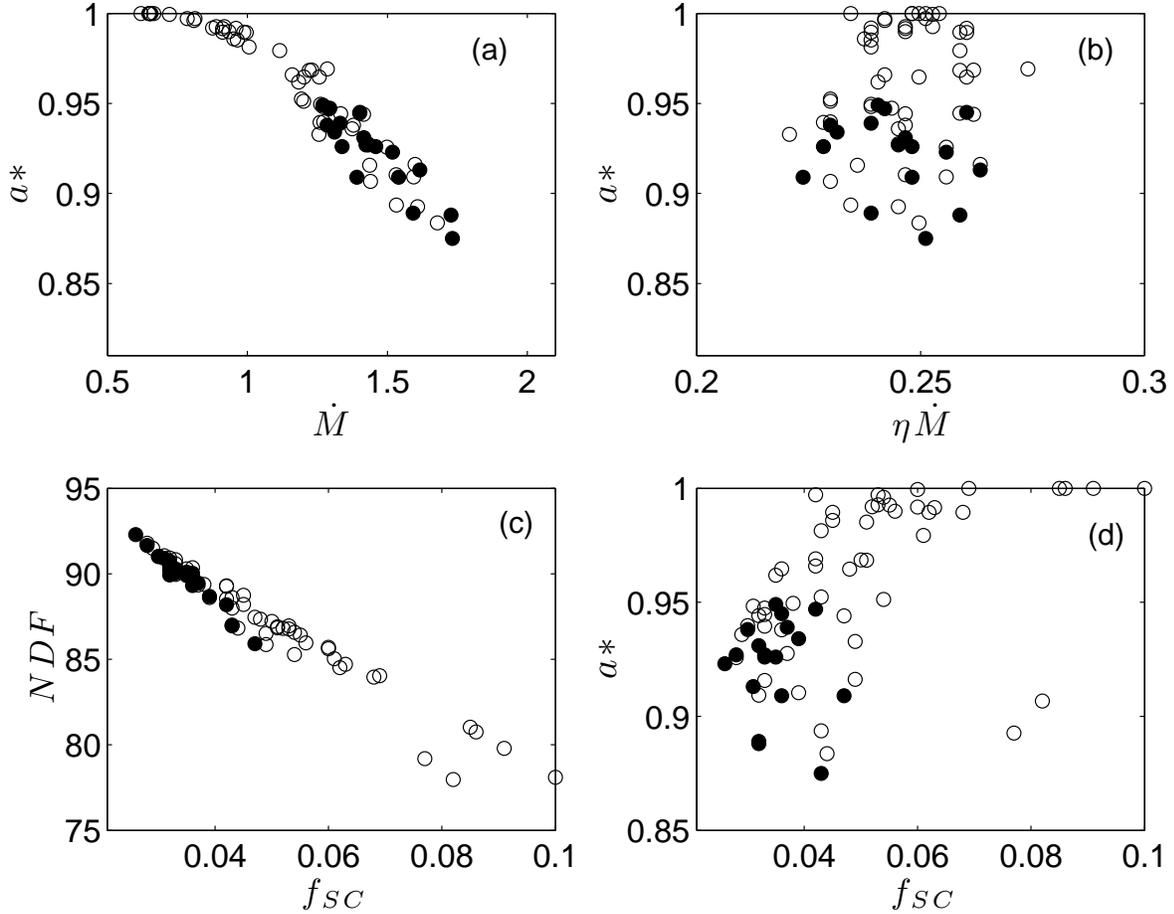}
\caption{Correlations between pairs of fitting parameters for our
  adopted model {\sc phabs(simpl}$\otimes${\sc kerrbb{\small 2}}) with
  $\Gamma$ fixed at 2.5.  Data for the 18 gold/silver observations are
  plotted as filled circles and for the 35 bronze observations as open
  circles. The mass accretion rate $\dot{M}$ is in units of
  $10^{18}~{\rm g~s^{-1}}$, and $\eta$ is the efficiency for
  converting rest mass into radiant energy.}
\label{rxte_correlation}
\end{figure}

\begin{figure}[ht]
\centering
\plotone{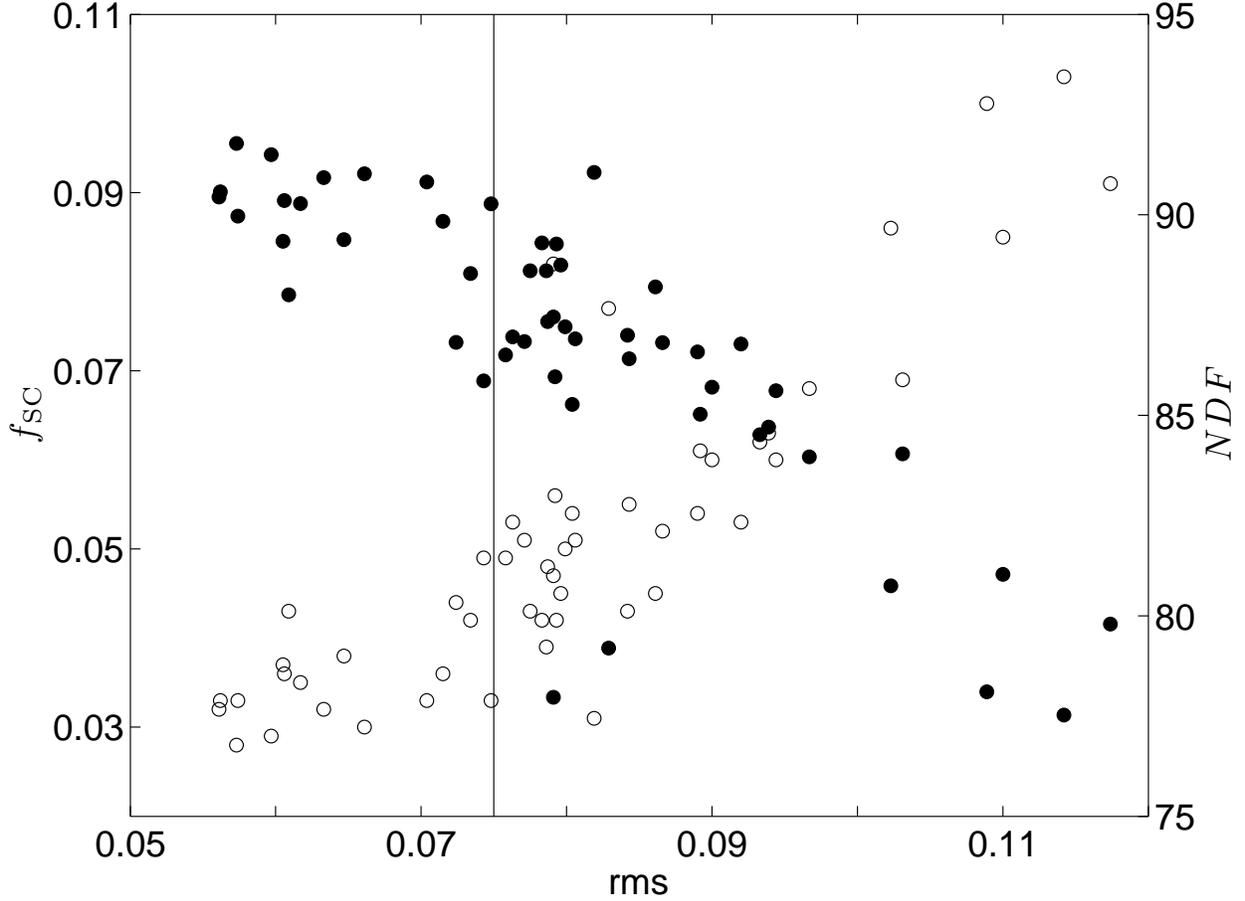}
\caption{The scattered fraction $f_{\rm SC}$ (open circles, scale on
  the left) and the naked disk fraction $NDF$ (filled circles, scale
  on the right) versus the rms power (0.1-10 Hz) for our standard
  model {\sc phabs(simpl}$\otimes${\sc kerrbb{\small 2}}) with
  $\Gamma$ fixed at 2.5.  The solid vertical line separates the data
  for our adopted sample of 18 gold/silver spectra (rms power $<
  0.075$) from that of the 35 bronze spectra.  We find very similar
  correlations for $\Gamma = 2.3$ and $\Gamma = 2.7$, i.e., the
  scattered fraction generally increases with increasing $\Gamma$
  while the disk fraction decreases.  This has the effect of shifting
  the plots up or down relative to the $\Gamma = 2.5$ plot shown
  here.}
\label{rxte_rms_sf}
\end{figure}

\newpage

\begin{figure}[ht]
\centering
\plotone{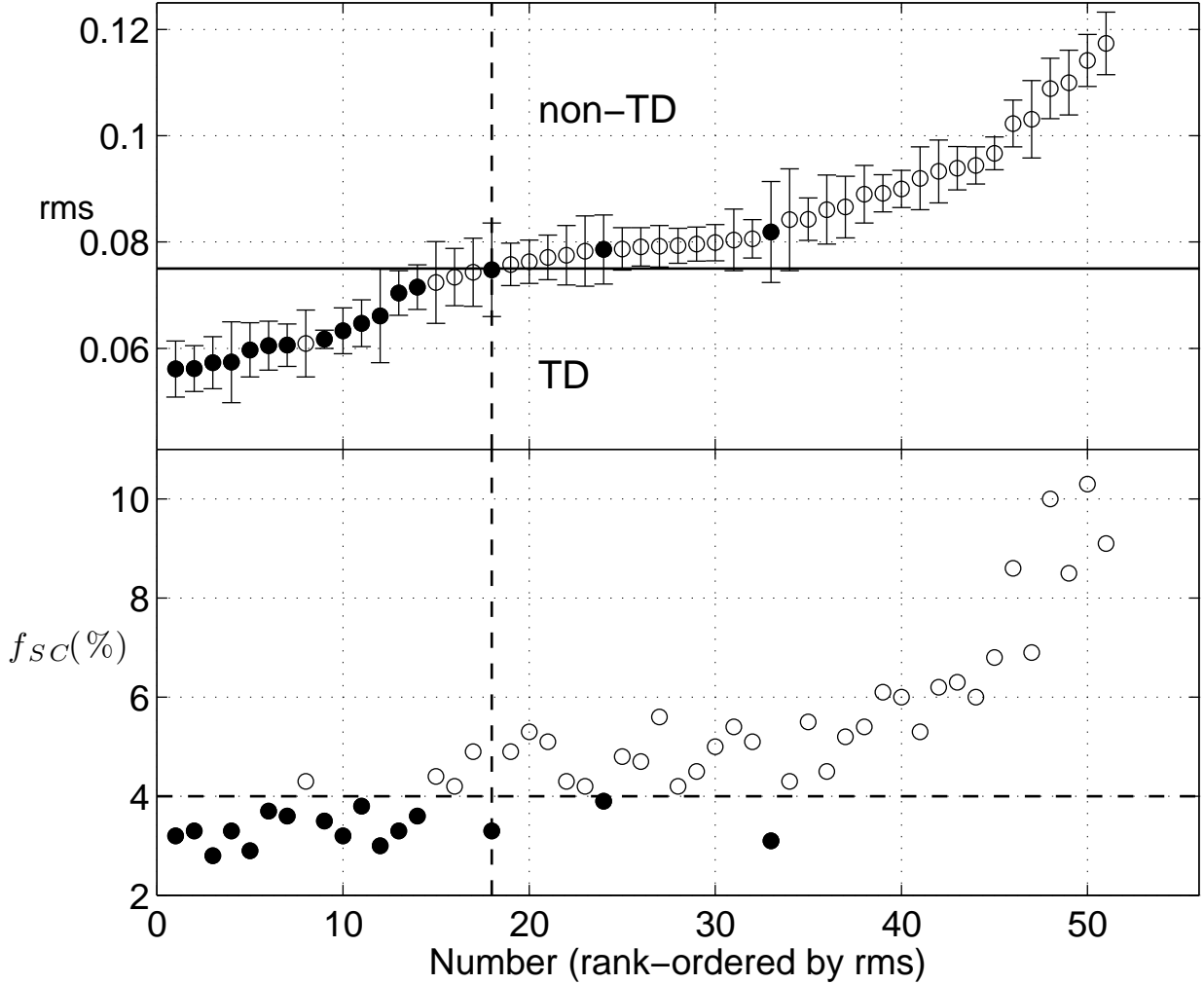}
\caption{ Exploration of the relationship of the classical TD rms
  criterion (rms power $< 0.075$) to the scattered fraction.  The
  spectra are plotted in order of increasing rms power.  {\bf Top:}
  The first 18 data points with rms power $< 0.075$ (indicated by the
  vertical dashed line and solid horizontal line) correspond to our
  gold/silver spectra and the remaining 35 are the bronze spectra.
  {\bf Bottom:} The horizontal dashed line indicates the
  scattered-fraction cut line at 4\% that corresponds to the rms power
  = 0.075 line (see text).  The spectra with $f_{\rm SC}< 4\%$ are
  indicated by filled circles in the top panel.  As shown, there is a
  very good correspondence between favored spectra with low rms power
  and those with low scattered fraction.}
\label{rxte_rms_df_corr}
\end{figure}

\newpage

\begin{figure}[ht]
\centering
\plotone{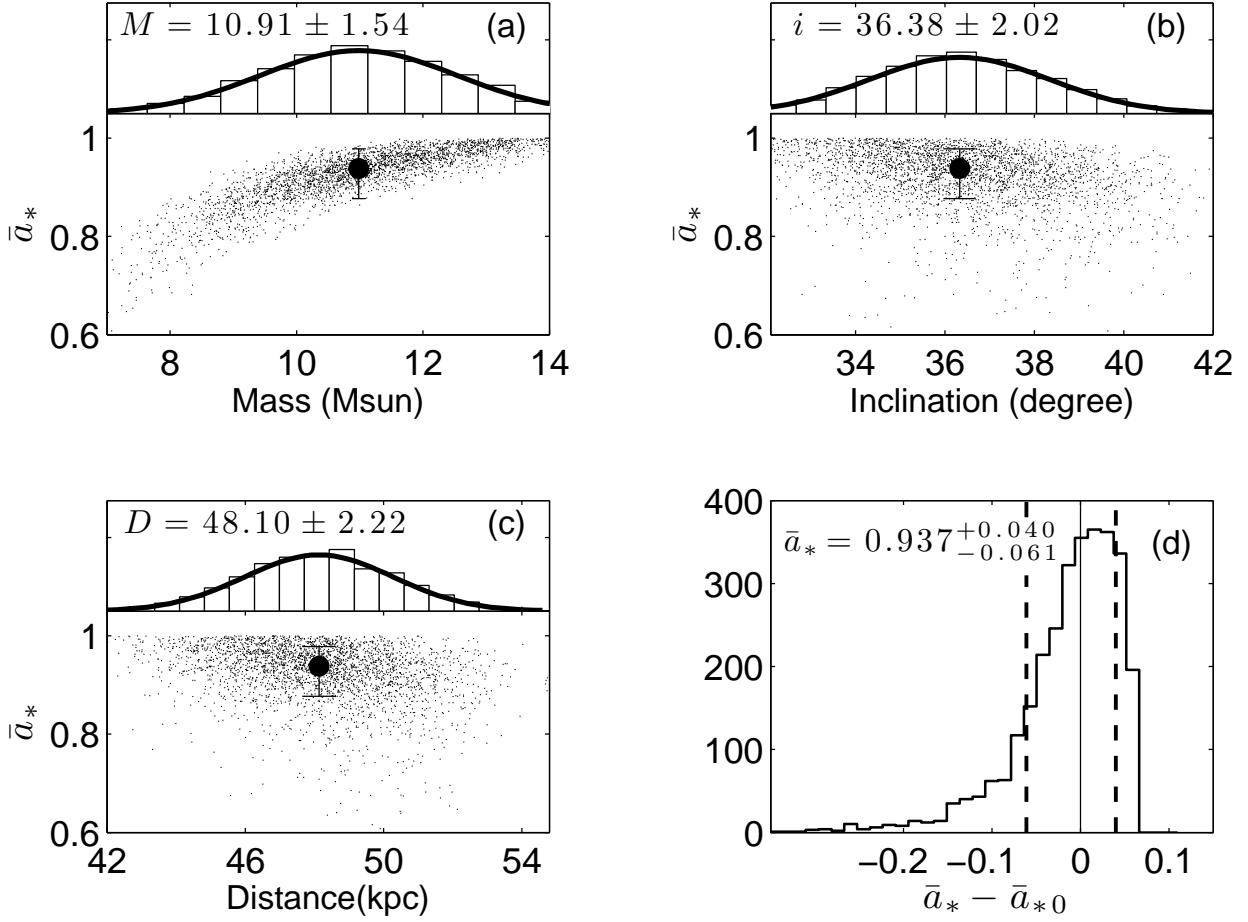}
\caption{Effect on the spin parameter $a_*$ of varying the input
  parameters $M$, $i$, and $D$ for the viscosity parameter
  $\alpha=0.01$.  ({\bf a}) The upper panel shows a normal
  distribution for the black hole mass and the lower panel shows $a_*$
  versus mass $M$ for 3000 sets of parameters drawn at random.  The
  central filled circle indicates our estimate of the spin
  ($\bar{a}_{*0} = 0.937_{-0.061}^{+0.040}$) obtained from these
  simulations.  ({\bf b,c}) Same as panel $a$ except now for the
  parameters of inclination angle and distance, respectively.  That
  the uncertainty in the mass is the dominant source of observational
  error is indicated in panel $a$ by the significant correlation
  between $a_*$ and $M$.  ({\bf d}) Histogram of spin displacements
  for 3000 parameter sets.  The vertical solid line indicates the
  median value of the spin in the simulated sample, $\bar{a}_{*0} =
  0.937$.  The two dashed lines enclose 68.27\% ($1\sigma$) of the
  spin values centered on the solid line and imply an observational
  uncertainty of (+0.040, -0.061).}
\label{rxte_error_analysis}
\end{figure}

\begin{figure}[ht]
\centering
\plotone{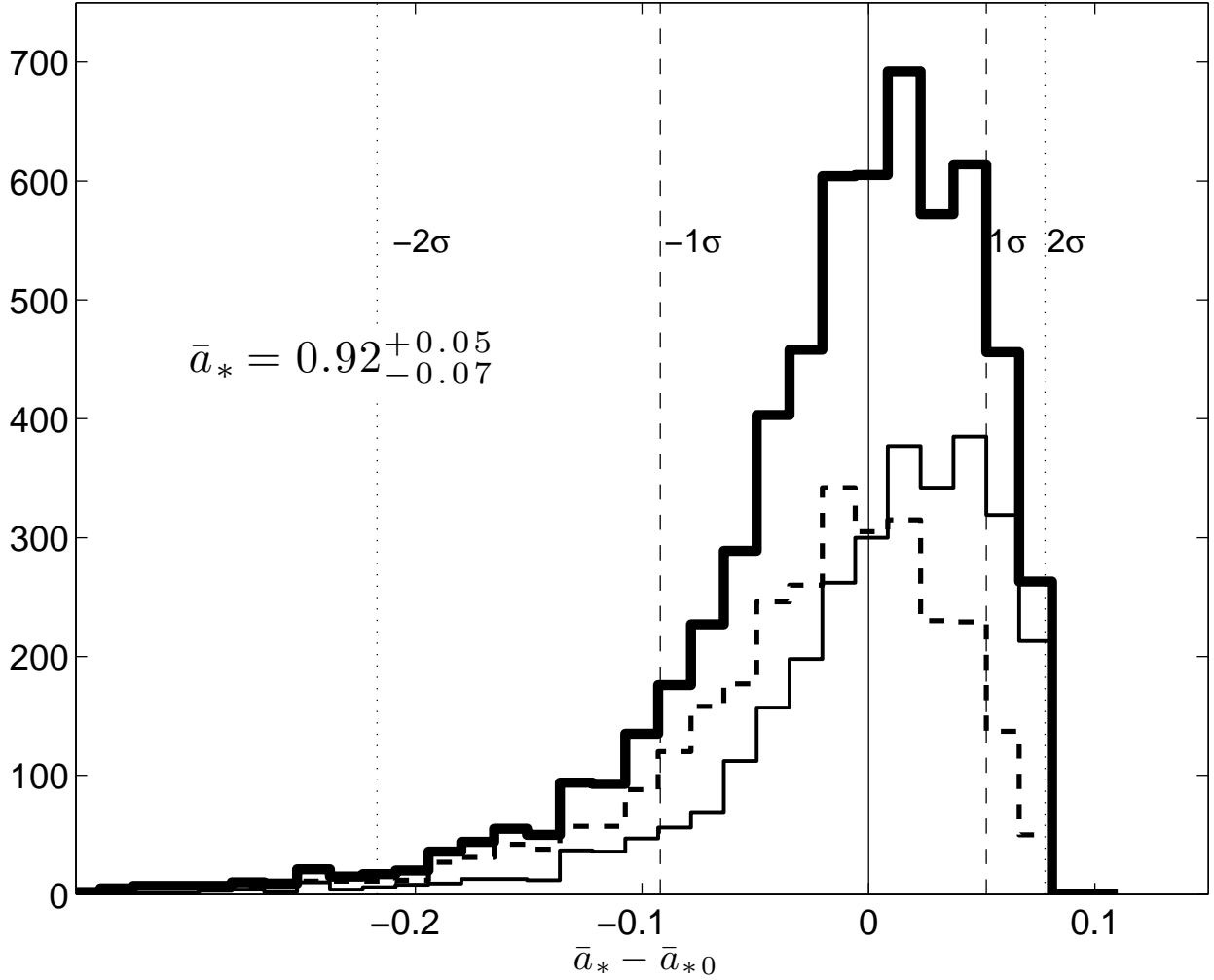}
\caption{Combined error analysis that considers both of our fiducial
  values for the viscosity parameter.  The thin solid line is for
  $\alpha=0.01$ and the dashed line is for $\alpha=0.1$.  The sum of
  these two smaller histograms forms the large histogram (thick solid
  line), which shows the spin displacements for a total of 6000
  parameter sets.  The vertical solid line indicates the median value
  of the spin determined by these simulations: $\bar{a}_{*0} = 0.92$.
  The two dashed lines enclose 68.27\% ($1\sigma$) of the spin values
  centered on the solid line and imply an observational and
  model-parameter uncertainty of (+0.05, -0.07). The two dotted lines
  enclose 95.45\% ($2\sigma$) of the spin values and imply an
  uncertainty of (+0.06, -0.18).}
\label{different_viscosity}
\end{figure}

\newpage

\begin{figure}[ht]
\centering
\plotone{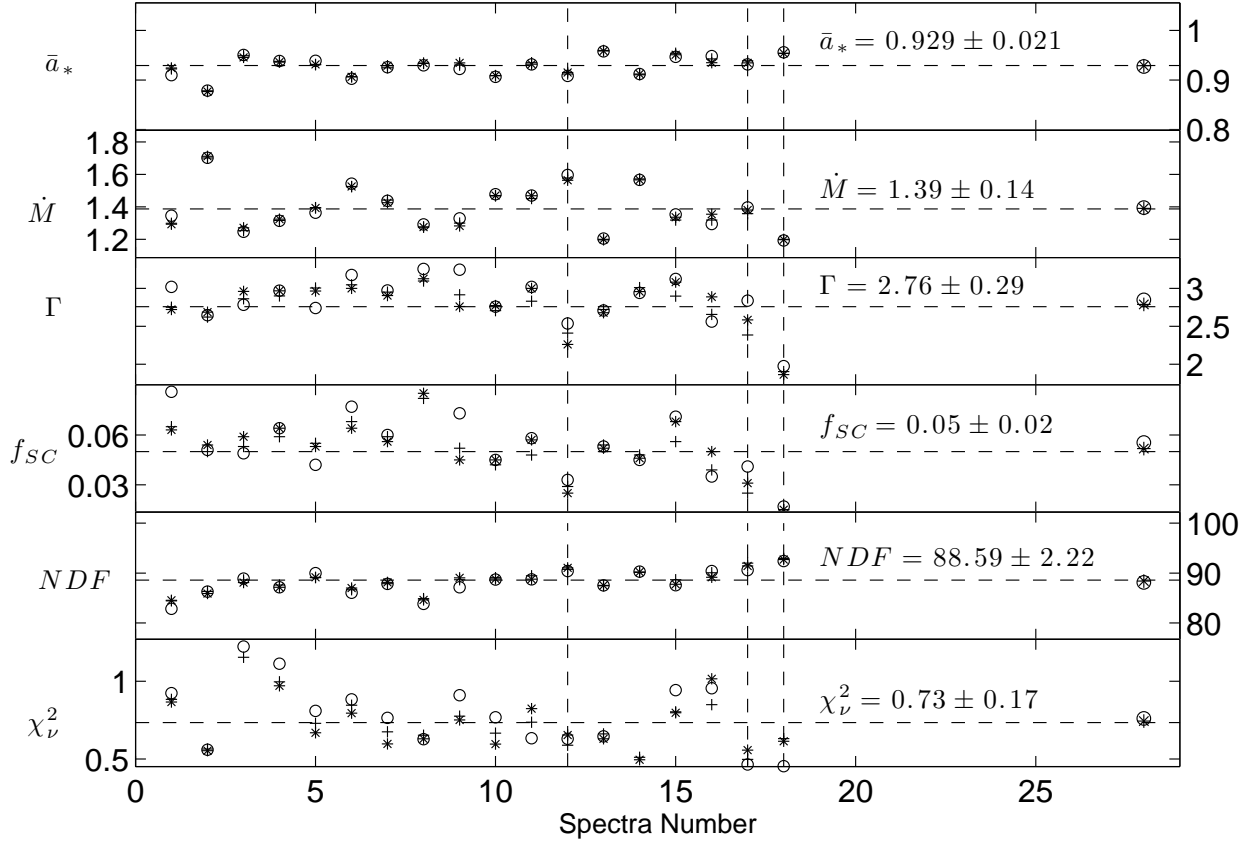}
\caption{Evidence that the nearby pulsar PSR B0540-69 does not
  significantly contaminate the 18 gold/silver spectra (see Appendix
  B).  In making this test, we used the model {\sc
    phabs(simpl}$\otimes${\sc kerrbb{\small 2}}) and allowed the
  power-law index $\Gamma$ to be a free parameter.  The top four
  panels show our standard fitting parameters that are defined in
  \S4.3 and the following panel shows the naked disk fraction $NDF$,
  which is defined in \S4.2. The bottom panel gives the reduced
  chi-square of the fit.  The fits were performed over three energy
  ranges: 2.5--15.0 keV (open circle), 2.5--20.0 keV (plus symbol) and
  2.5--25.0 keV (asterisk).  The mean value of each parameter for our
  nominal energy interval of 2.5--20.0 keV is indicated by a
  horizontal dashed line and is given directly above the dashed line
  inside each panel; the mean values for the other two energy
  intervals are indicated by the symbols plotted at the far right of
  the figure.  The vertical dashed lines indicate the three gold TD
  spectra (Nos.\ 12, 17 \& 18).}
\label{pulsar_effect}
\end{figure}

\end{document}